\documentclass{emulateapj}

\def\lapp{\ifmmode\stackrel{<}{_{\sim}}\else$\stackrel{<}{_{\sim}}$\fi}
\def\gapp{\ifmmode\stackrel{>}{_{\sim}}\else$\stackrel{>}{_{\sim}}$\fi}
\def\deg{$^\circ$\xspace}
\def\mss{$\mathrm{m/s^2}$\xspace}
\usepackage{graphicx}
\usepackage{xspace}
\usepackage{longtable}
\usepackage{dcolumn}
\newcolumntype{d}[1]{D{.}{.}{#1}}
\newcolumntype{X}[1]{D{-}{-}{#1}}

\newcommand{\sigproc}{\texttt{SIGPROC}\xspace}
\newcommand{\presto}{\texttt{PRESTO}\xspace}
\newcommand{\psrfits}{\texttt{PSRFITS}\xspace}
\newcommand{\rfifind}{\texttt{rfifind}\xspace}
\newcommand{\realfft}{\texttt{realfft}\xspace}
\newcommand{\prepsub}{\texttt{prepsubband}\xspace}
\newcommand{\accelsearch}{\texttt{accelsearch}\xspace}
\newcommand{\prepfold}{\texttt{prepfold}\xspace}
\newcommand{\spsearch}{\texttt{single\_pulse\_search.py}\xspace}
\newcommand{\ddplan}{\texttt{DDplan.py}\xspace}
\newcommand{\rednoise}{\texttt{rednoise}\xspace}
\newcommand{\nemodel}{\texttt{NE2001}\xspace}

\newcommand{\psrfitsutils}{\texttt{psrfits\_utils}\xspace}
\newcommand{\foldpsrfits}{\texttt{fold\_psrfits}\xspace}
\newcommand{\combinemocks}{\texttt{combine\_mocks}\xspace}
\newcommand{\fluxcal}{\texttt{fluxcal}\xspace}
\newcommand{\psrchive}{\texttt{psrchive}\xspace}
\newcommand{\psrpoppy}{\texttt{PsrPopPy}\xspace}
\newcommand{\snr}{$S/N$\xspace}
\newcommand{\snrs}{$S/N$s\xspace}
\newcommand{\sigfourier}{\ifmmode\sigma_F\else$\sigma_F$\fi\xspace}
\newcommand{\snrtime}{\ifmmode\left ( S/N \right )_T\else$\left( S/N \right )_T$\fi\xspace}
\newcommand{\guilly}{Guillimin\xspace}
\newcommand{\dmunit}{$\mathrm{pc\,cm^{-3}}$\xspace}
\newcommand{\sqdeg}{sq.~deg.\xspace}
\newcommand{\sigmaloc}{\ifmmode\sigma_\mathrm{loc}\else$\sigma_\mathrm{loc}$\fi\xspace}

\newcommand{\app}[1]{$\sim$\,#1}
\newcommand{\sect}{\S}
\newcommand{\sects}{\S\S}
\newcommand{\hdr}[1]{\multicolumn{1}{c}{#1}}
\newcommand{\nts}[1]{#1}

\newcommand{\removed}[1]{}

\begin{document}

\title{Arecibo Pulsar Survey Using ALFA. IV. Mock Spectrometer Data Analysis, Survey Sensitivity, and the Discovery of 41 Pulsars}

\author{
P.~Lazarus\altaffilmark{1},
A.~Brazier\altaffilmark{2,3},
J.~W.~T.~Hessels\altaffilmark{4,5},
C.~Karako-Argaman\altaffilmark{6},
V.~M.~Kaspi\altaffilmark{6},
R.~Lynch\altaffilmark{7,6},
E.~Madsen\altaffilmark{6},
C.~Patel\altaffilmark{6},
S.~M.~Ransom\altaffilmark{8},
P.~Scholz\altaffilmark{6},
J.~Swiggum\altaffilmark{7},
W.~W.~Zhu\altaffilmark{1},
B.~Allen\altaffilmark{9,10,11},
S.~Bogdanov\altaffilmark{12},
F.~Camilo\altaffilmark{12},
F.~Cardoso\altaffilmark{7},
S.~Chatterjee\altaffilmark{2},
J.~M.~Cordes\altaffilmark{2},
F.~Crawford\altaffilmark{13},
J.~S.~Deneva\altaffilmark{14},
R.~Ferdman\altaffilmark{6},
P.~C.~C.~Freire\altaffilmark{1},
F.~A.~Jenet\altaffilmark{15}, 
B.~Knispel\altaffilmark{9},  
K.~J.~Lee\altaffilmark{16},
J.~van~Leeuwen\altaffilmark{4,5}, 
D.~R.~Lorimer\altaffilmark{7},
A.~G.~Lyne\altaffilmark{17}, 
M.~A.~McLaughlin\altaffilmark{7},
X.~Siemens\altaffilmark{11}, 
L.~G.~Spitler\altaffilmark{1},
I.~H.~Stairs\altaffilmark{18}, 
K.~Stovall\altaffilmark{19}, 
A.~Venkataraman\altaffilmark{20}
}

\altaffiltext{1}{Max-Planck-Institut f\"ur Radioastronomie, Auf dem H\"ugel 69, 53121 
                 Bonn, Germany; plazarus@mpifr-bonn.mpg.de}
\altaffiltext{2}{Dept. of Astronomy, Cornell Univ., Ithaca, NY 14853, USA} 
\altaffiltext{3}{Center for Advanced Computing, Cornell Univ., Ithaca, NY 14853, USA}
\altaffiltext{4}{ASTRON, the Netherlands Institute for Radio Astronomy, Postbus 2, 7990 AA, 
                 Dwingeloo, The Netherlands} 
\altaffiltext{5}{Anton Pannekoek Institute for Astronomy, Univ. of Amsterdam, Science 
                 Park 904, 1098 XH Amsterdam, The Netherlands}
\altaffiltext{6}{Dept.~of Physics, McGill Univ., Montreal, QC H3A 2T8, Canada}
\altaffiltext{7}{Dept.~of Physics, West Virginia Univ., Morgantown, WV 26506, USA} 
\altaffiltext{8}{NRAO, Charlottesville, VA 22903, USA} 
\altaffiltext{9}{Max-Planck-Institut f\"ur Gravitationsphysik, D-30167 Hannover, Germany}
\altaffiltext{10}{Leibniz Universit{\"a}t Hannover, D-30167 Hannover, Germany}
\altaffiltext{11}{Physics Dept., Univ. of Wisconsin - Milwaukee, Milwaukee WI 53211, USA}
\altaffiltext{12}{Columbia Astrophysics Laboratory, Columbia Univ., New York, NY 10027, USA} 
\altaffiltext{13}{Dept.~of Physics and Astronomy, Franklin and Marshall College, Lancaster, 
                  PA 17604-3003, USA} 
\altaffiltext{14}{Naval Research Laboratory, 4555 Overlook Ave. SW, Washington, DC 20375, USA}
\altaffiltext{15}{Center for Gravitational Wave Astronomy, Univ.~of Texas - Brownsville, 
                  TX 78520, USA} 
\altaffiltext{16}{Kavli Institute for Astronomy and Astrophysics, Peking Univ., Beijing 
                  100871, P.R. China}
\altaffiltext{17}{Jodrell Bank Centre for Astrophysics, Univ.~of Manchester, Manchester, 
                  M13 9PL, UK} 
\altaffiltext{18}{Dept.~of Physics and Astronomy, Univ.~of British Columbia, Vancouver, 
                  BC V6T 1Z1, Canada} 
\altaffiltext{19}{Dept.~of Physics and Astronomy, Univ.~of New Mexico, NM 87131, USA}
\altaffiltext{20}{Arecibo Observatory, HC3 Box 53995, Arecibo, PR 00612}

\begin{abstract}
The on-going PALFA survey at the Arecibo Observatory began in 2004 and is
searching for radio pulsars in the Galactic plane at 1.4~GHz.  Observations
since 2009 have been made with new wider-bandwidth spectrometers than were
previously employed in this survey. A new data reduction pipeline has been in
place since mid-2011 which consists of standard methods using dedispersion,
searches for accelerated periodic sources, and search for single pulses, as
well as new interference-excision strategies and candidate selection
heuristics.  This pipeline has been used to discover 41 pulsars, including 8
millisecond pulsars (MSPs; $P < 10$\,ms), bringing the PALFA survey's discovery
totals to 145 pulsars, including 17 MSPs, and one Fast Radio Burst (FRB). The
pipeline presented here has also re-detected 188 
previously known pulsars including 60 found in PALFA data by re-analyzing
observations previously searched by other pipelines.  A comprehensive
description of the survey sensitivity, including the effect of interference and
red noise, has been determined using synthetic pulsar signals with various
parameters and amplitudes injected into real survey observations and
subsequently recovered with the data reduction pipeline.  We have
confirmed that the PALFA survey achieves the sensitivity to MSPs predicted by
theoretical models. However, we also find that compared to theoretical
survey sensitivity models commonly used there is a degradation in
sensitivity to pulsars with periods $P\gapp100$\,ms that gradually becomes up
to a factor of \app{10} worse for $P>4$\,s at $\mathrm{DM} < 150$\,\dmunit.
This degradation of sensitivity at long periods is largely due to red noise.
We find that \nts{$35 \pm 3$\,\%} of pulsars are missed despite being
bright enough to be detected in the absence of red noise.  This reduced
sensitivity could have implications on estimates of the number of long-period
pulsars in the Galaxy.
\end{abstract}


\section{Introduction}
\label{sec:Introduction}
Pulsars are rapidly rotating, highly magnetized neutron stars, the remnants of
massive stars after their death in supernova explosions. They are extremely
valuable astronomical tools with many physical applications that have been used
to, for example,
constrain the equation of state of ultra-dense matter
\citep[e.g.][]{hrs+06,dpr+10}, test relativistic gravity
\citep[e.g][]{ksm+06,afw+13}, probe plasma physics within the magnetosphere
\citep[e.g.][]{hkwe03,klo+06,lhk+10,hhk+13}, and gain a better understanding of the
complete radio pulsar population \citep[e.g.][]{fk06}.  Certain individual
pulsar systems are especially well suited to studying these areas of
astrophysics, and thus continued pulsar surveys to find these rare objects
remain a major scientific driver in the field. 

Radio pulsars are found primarily in non-targeted, wide-area surveys such as the
Pulsar-ALFA (PALFA) survey at 1.4\,GHz, which began in 2004 \citep{cfl+06}.
PALFA observations use the 7-beam Arecibo L-band Feed Array (ALFA) receiver
of the Arecibo Observatory William E. Gordon 305-m Telescope and focus on
the Galactic plane ($|b| < 5$\deg) in the two regions visible with Arecibo,
namely the ``inner Galaxy'' region (32\deg \lapp~$l$ \lapp~77\deg), and the
``outer Galaxy'' region (168\deg \lapp~$l$ \lapp~214\deg). 

For the first 5 years, PALFA survey observations were made using the Wideband
Arecibo Pulsar Processor (WAPP), a 3-level auto-correlation spectrometer with
100~MHz of bandwidth \citep{dsh00}.  Since 2009, the Mock
spectrometer\footnote{http://www.naic.edu/$\sim$astro/mock.shtml}, a 16-bit
poly-phase filterbank with 322~MHz of bandwidth, has replaced the WAPP
spectrometer as the data-recorder of the PALFA survey. The increased bandwidth,
poly-phase filterbank design, and increased bit-depth of the Mock spectrometer
have increased the sensitivity and robustness to interference of the PALFA
survey. For this reason, we are re-observing regions of the sky
previously observed with the WAPP spectrometers.

The PALFA consortium currently employs two independent full-resolution data
analysis pipelines. The Einstein@Home-based pipeline
(E@H)\footnote{http://einstein.phys.uwm.edu/} has already been described by
\citet{akc+13}: this pipeline derives its computational power by aggregating
the spare cycles of a global network of PCs and mobile devices using the BOINC
platform, and is also searching data from the PALFA survey for pulsars. In this
work we describe the pipeline based on the PRESTO suite of pulsar search
programs\footnote{http://www.cv.nrao.edu/$\sim$sransom/presto/} \citep{ran01}.
In addition to these pipelines, we also employ a reduced-resolution ``Quicklook''
pipeline, which is run on-site at Arecibo shortly after observing sessions are
complete and which enables a more rapid discovery and confirmation of strong
pulsars \citep{sto13}.  

As of March 2015, there have been 145 pulsars discovered in WAPP and Mock
spectrometer observations with the various PALFA data analysis pipelines. This
is already a sizable increase on the known sample of 258 Galactic radio pulsars
in the full survey region found in other searches\footnote{As listed in the
ATNF Pulsar Catalogue: http://www.atnf.csiro.au/research/pulsar/psrcat
\citep{mhth05}}.


The relatively high observing frequency and unparalleled sensitivity of Arecibo,
coupled with the high time and frequency resolution of PALFA ($\tau_\mathrm{samp}
\simeq\!65.5\,\mu$s and $\Delta f_\mathrm{chan} \simeq\!336$\,kHz, respectively) make it
particularly well suited for detecting millisecond pulsars (MSPs) deep in the
plane of the Galaxy, such as the distant MSPs reported by \citet{csl+12} and
\citet{skl+15}, the highly eccentric MSP PSR~J1903+0327
\citep{crl+08}, and faint, young pulsars \citep[e.g.][]{hng+08}. 
The huge instantaneous sensitivity of Arecibo enables short integration times,
which has been helpful in detecting relativistic binaries \citep[e.g.
PSR~J1906+0746;][]{lsf+06} by reducing the deleterious effect of time-varying
Doppler shifts of binary pulsars. The PALFA survey has also proven successful
at detecting transient astronomical signals. For example, the survey has led to
the discovery of several Rotating Radio Transient pulsars
\citep[RRATs;][]{dcm+09}, as well as FRB~121102, the first Fast Radio Burst
(FRB) detected with a telescope other than the Parkes Radio Telescope
\citep{sch+14}.

While PALFA is the most sensitive large-scale survey for radio pulsars ever
conducted, it is not the only on-going radio pulsar survey. Other major
surveys are the HTRU-S \citep{kjs+10}, HTRU-N \citep{bck+13},
and SPAN512 \citep{dcc+13} surveys at \app{1.4\,GHz}, the
GBNCC \citep{slr+14} and AO327 drift \citep{dsm+13} surveys at \app{350\,MHz},
and the LOFAR surveys \citep{cvh+14} at \app{150\,MHz}.

The underlying distributions of the pulsar population can
be estimated using simulation techniques \citep[e.g.][]{lfl+06,fk06}. The large
sample of pulsars found in non-targeted surveys are essential for these
simulations. However, for population analyses to be done accurately, the
selection biases of each survey must be taken into account. While the
sensitivity of pulsar search algorithms is reasonably well understood, the
effect of radio frequency interference (RFI) on pulsar detectability has not been
previously studied in detail.

This paper reports on the current state of PALFA's primary search pipeline,
its discoveries, and its sensitivity. The rest of the article is
organized as follows:
the observing set-up is summarized in \sect~\ref{sec:observations}. The details
of the PALFA \presto-based pipeline are described in \sect~\ref{sec:analysis}.
\sect~\ref{sec:results} reports basic parameters of the pulsars found with the
pipeline, and \sect~\ref{sec:injected} details how the survey sensitivity is
determined, including a technique involving injecting synthetic pulsars into
the data. These accurate sensitivity limits are used to improve upon population
synthesis analyses in \sect  \ref{sec:popsynth}. The broader implications of the
accurate determination of the survey sensitivity are presented in
\sect~\ref{sec:discussion} before the paper is summarized in
\sect~\ref{sec:conclusion}.

\section{Observations}
\label{sec:observations}

The PALFA survey observations have been restricted to the two regions of the
Galactic plane ($|b| < 5$\deg) visible from the Arecibo observatory, the inner
Galaxy (32\deg \lapp~$l$ \lapp~77\deg), and the outer Galaxy (168\deg \lapp~$l$
\lapp~214\deg). Integration times are 268~s and 180~s for inner and outer
Galaxy observations, respectively.

To optimize the use of telescope resources, the PALFA survey operates in tandem
with other compatible projects using the ALFA 7-beam receiver. In particular,
we have reciprocal data-sharing agreements with collaborations that search for
galaxies in the optically obscured (``zone of avoidance'') directions through the
Milky Way 
\citep{hsb+10} and recombination-line studies of ionized gas in the Milky Way 
\citep{lmt+13}. The PALFA project leads inner Galaxy observing sessions,
whereas our partners lead outer Galaxy sessions.

\begin{figure*}[t]
        \includegraphics[width=\textwidth,angle=0]{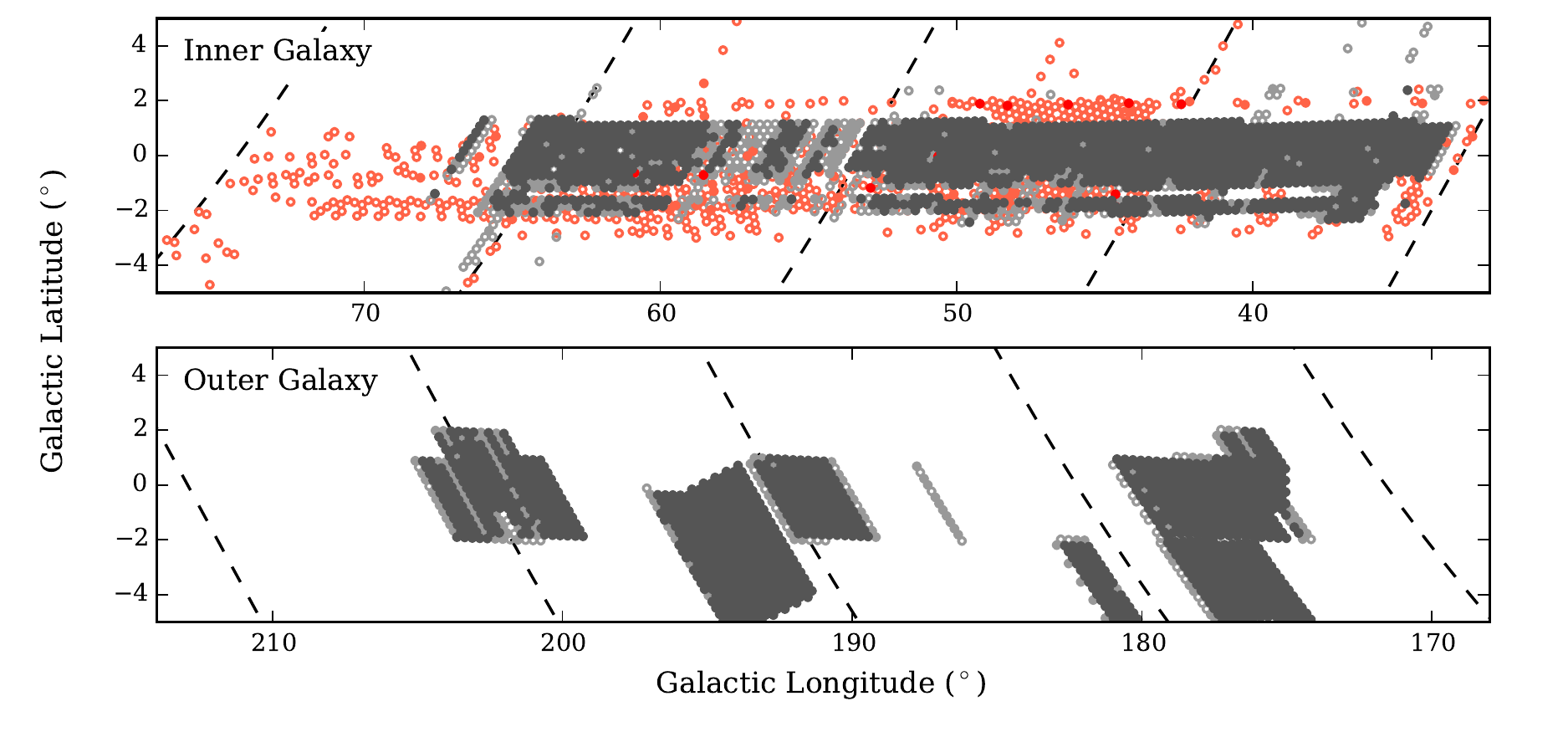}
        \caption{Sky map showing the locations of PALFA observations with the
        Mock spectrometers, which began in 2009, for the inner and outer Galaxy
        regions. Each position plotted represents the center of the
        3-pointing set required to densely sample the area.  Positions that
        have been only sparsely observed (i.e. 1 of 3 pointing positions
        observed) are indicated with un-filled circles. Positions with 2 of 3
        pointings observed are indicated with a light-colored filled circle.
        Positions that have been densely observed (i.e. all 3 pointing
        positions observed) are indicated with dark-colored filled circles. Red
        indicates observations made prior to adjusting our pointing grid at the
        request of our commensal partners. As a result, some of the sky
        area covered in early Mock observations has not been re-observed
        using the Mocks and the current commensal pointing grid.
    \label{fig:skymap}}
\end{figure*}

For the inner Galaxy region, the pointing strategy has prioritized 
observations of the $|b| < 2$\deg region before densely sampling the Galactic
plane at larger Galactic latitudes. To densely cover a patch of sky out to the
ALFA beam FWHM, three interleaved ALFA pointings are required \citep[see][for
more details]{cfl+06}.  In contrast, our commensal partners have focused
outer Galaxy observations in order to densely sample particular Galactic
longitude/latitude ranges. A sky map showing the observed pointing positions
can be found in Figure~\ref{fig:skymap}.

Observations conducted with ALFA have a bandwidth of 322\,MHz centered at
1375\,MHz. Each of the seven ALFA beams is split into two overlapping 172-MHz
sub-bands and processed independently by the Mock
spectrometers\footnote{http://www.naic.edu/$\sim$astro/mock.shtml}. The
sub-bands are divided into 512 channels, each sampled every $\sim$65.5 $\mu$s.
The observing parameters are summarized in Table \ref{tab:obs_setup}. The data
are recorded to disk in 16-bit search-mode PSRFITS format \citep{hvm04}. 

PALFA survey data have been recorded with the Mock spectrometers since 2009.
Although, in 2011 our pointing grid was altered slightly to accommodate our
commensal partners. This required some sky positions to be re-observed. 
Prior to 2009, survey observations were recorded with the Wideband Arecibo
Pulsar Processors \citep[WAPPs; see][]{dsh00,cfl+06}. The two data recording
systems were run in parallel during 2009 to check the consistency and quality
of the Mock spectrometer data.

An unpulsed calibration diode is fired during the first (or sometimes last)
5--10\,s of our integration. While this is primarily used by our partners, we
have found the diode signals useful in calibrating observations for our
sensitivity analysis (see \sect~\ref{sec:sensitivity}). The calibration signal
is removed from the data prior to searching (see
\sect~\ref{sec:pipeline-preproc}).

The original 16-bit Mock data files are compressed to have 4 bits per sample. These
smaller data files are more efficient to ship and analyze thanks to reduced
disk-space requirements. The 4-bit data files utilize the
\textit{scales} and \textit{offsets} fields of the PSRFITS format to retain
information about the bandpass shape despite the reduced dynamic range. The
scales and offsets are computed and stored for every 1-s sub-integration. 
This reduction of bit-depth results in a total loss of only a few percent
in the \snr of pulsar signals.

The converted 4-bit PSRFITS data files are copied to hard disks, and
couriered from Arecibo to Cornell University where they are archived at the
Cornell University Center for Advanced Computing. Meta-data about each
observation, parsed from the telescope logs and the file headers, are stored
in a dedicated database.

As of 2014 November, 
a total of 87689 beams of Mock spectrometer data have been archived. The
break-down of observed, archived and analyzed sky positions for the two survey
regions is shown in Table \ref{tab:data}.

PALFA observations more than one year old are publicly available. Small
quantities of data can be requested via the
web\footnote{http://arecibo.tc.cornell.edu/PalfaDataPublic}. Access to larger
amounts of data is also possible, but must be coordinated with the
collaboration because of logistics. 

Additional details about the data management logistics and data preparation are
in \sects~\ref{sec:pipeline-logistics} and \ref{sec:pipeline-preproc}.

\section{Pulsar and Transient Search Pipeline}
\label{sec:analysis}
\label{sec:pipeline}

The \presto-based pipeline has been used to search PALFA observations taken
with the Mock spectrometers since mid-2011 for radio pulsars and transients.
All processing is done using the \guilly supercomputer of McGill University's
High Performance Computing center\footnote{http://www.hpc.mcgill.ca/}.

While the pipeline described here was designed specifically for the PALFA
survey, it is sufficiently flexible to serve as a base for the data reduction
pipeline of other surveys. For example, the SPAN512 survey being undertaken at
the Nan\c{c}ay Radio Telescope uses a version of the PALFA \presto pipeline
described here tuned to their specific needs \citep{dcc+13}. The PALFA pipeline
source code is publicly available online\footnote{https://github.com/plazar/pipeline2.0}.

Since the analysis began with the pipeline, there have been several major
improvements, primarily focusing on ameliorating its robustness in the presence of
RFI (\sect~\ref{sec:pipeline-rfi}), as well as post-processing algorithms for
identifying the best pulsar candidates (\sect~\ref{sec:pipeline-post}). The
PALFA consortium is constantly monitoring the performance of the pipeline and
the RFI environment at Arecibo (as described later, RFI is one of the major
challenges), and looking for ways to further improve the analysis.  Here we
report on the state of the software as of early-2015.  

The pipeline overview presented here is grouped into logical
components. 
In \sect~\ref{sec:pipeline-logistics} we outline the significant
data tracking and processing logistics required to automate the analysis.
In \sect~\ref{sec:pipeline-preproc} we detail the data file preparation
required before searching an observation.
In \sect~\ref{sec:pipeline-search} we describe the techniques used to search for
periodic and impulsive pulsar signals. 
In \sect~\ref{sec:pipeline-rfi} we summarize the various complementary stages of
RFI identification and mitigation.
Finally, in \sects~\ref{sec:pipeline-post} and \ref{sec:cyberska} we outline the
tools used to help select and view pulsar candidates, as well as other on-line
collaborative facilities used by the PALFA consortium.

\begin{figure*}[t]
        \includegraphics[width=\textwidth]{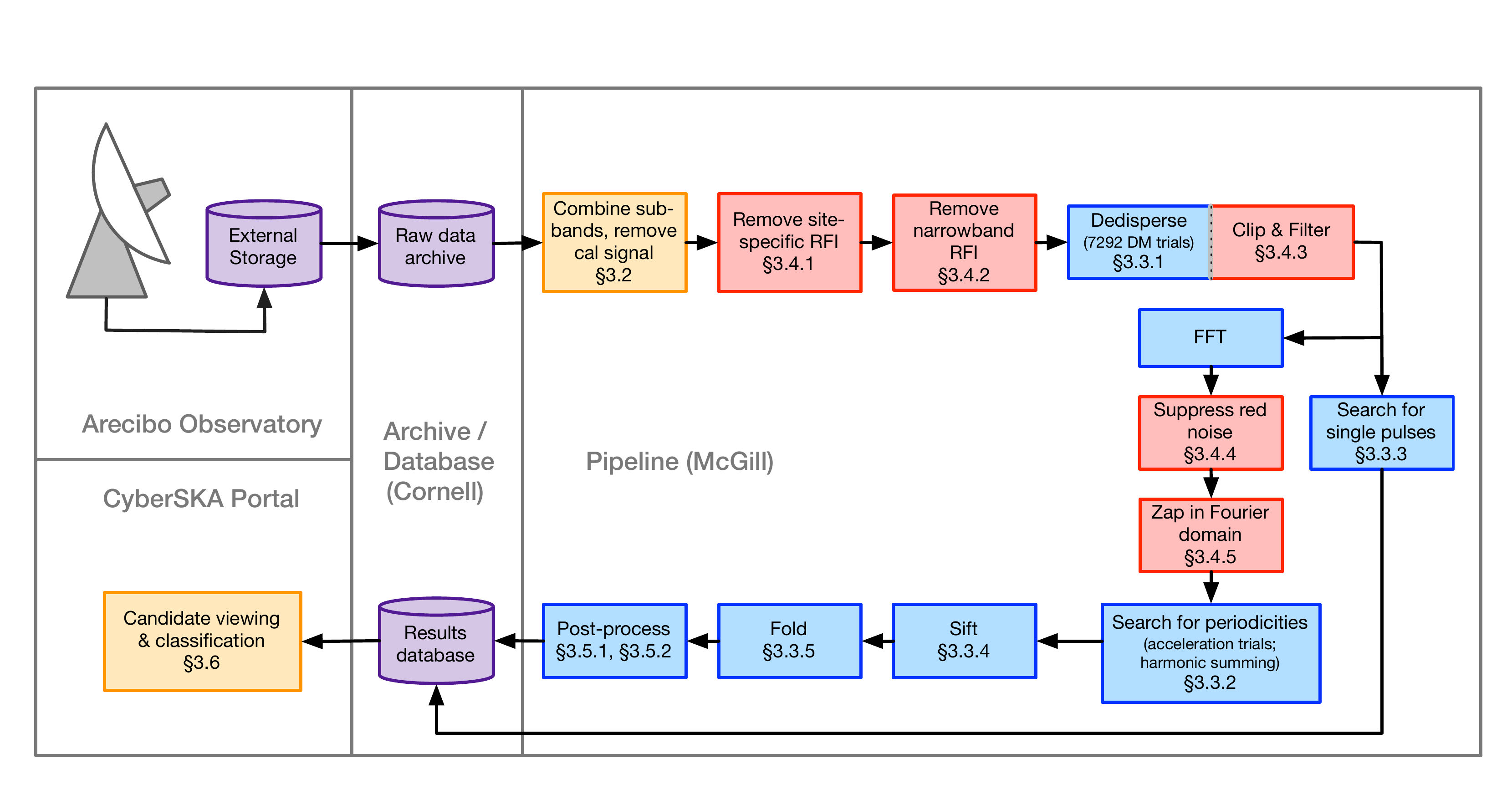}
        \caption{
            An overview of the PALFA survey's \presto-based pipeline. The
            color of each element reflects the category of the step: searching
            is blue; RFI mitigation is red; data storage and databases are
            purple; miscellaneous processes are yellow. Additional details about
            each pipeline stage can be found in the sections listed in each box.
    \label{fig:flowchart}}
\end{figure*}

Figure \ref{fig:flowchart} shows a flowchart summarizing the stages of the pipeline.

\subsection{Logistics}
\label{sec:pipeline-logistics}

The PALFA search pipeline is designed to be almost entirely automated. This
includes the logistics of data management required to maintain the analysis of
$\sim1000$ observations on the \guilly supercomputer at any given time. This
is accomplished with a \textit{job-tracker} database that maintains the status
of processes that are downloading raw data, reducing data, and uploading results.

The pipeline is configured to continually request and download raw data that
have not been processed and delete the local copies of files that have been
successfully analyzed. Data files are copied to McGill via FTP from the Cornell
University Center for Advanced Computing (CAC). The multi-threaded data
transfers from the CAC to McGill are sufficiently fast to maintain 1000--2000
jobs running simultaneously.

When the transfer of an observation is complete, job entries are created in the
pipeline's job-tracker database. As compute resources become available, jobs
are automatically submitted to the super-computer's queue.

When jobs terminate, the pipeline checks for results and errors. Failed jobs are
automatically re-submitted up to three times to allow for occasional hiccups of
the \guilly task management system, or processing node glitches. If all three
processing attempts result in failure, the observation is flagged to be dealt
with manually. Observations that are salvageable are re-processed after fixes
are applied. The positions of un-salvageable observations are re-inserted into
the observing schedule, along with those from observations severely
contaminated with RFI. Observations may be un-salvageable if they are aborted
scans, contain malformed metadata, or their files have become corrupted.
Only \app{0.15}\% of all observations have data files that cannot be searched,
and only \app{4.5}\% of all observations are flagged to be re-observed due to
excessive RFI.

The results from successfully processed jobs are parsed and uploaded to a
database at the CAC, and the local copies of the data files are removed to
liberate disk space enabling more observations to be requested, downloaded, and
analyzed.

The inspection of uploaded results is done with the aid of a web-application
(see \sect \ref{sec:cyberska}).

\subsection{Pre-processing}
\label{sec:pipeline-preproc}

Before analyzing the data for astrophysical signals, the two Mock sub-bands
must be combined into a single \psrfits file. Each of the two Mock data files
have 512 frequency channels, 66 of which are overlapping with the other file.
For each sub-integration of the observation, the 478 low-frequency channels
from the bottom sub-band and the 480 high-frequency channels from the top
sub-band are extracted, concatenated together -- along with two extra, empty
frequency channels -- for each sample, and written into a new full-band data
file, consisting of 960 channels. The choice to discard part of both bands was
made in order to mitigate the effect of the reduced sensitivity at the
extremities of the Mock sub-bands, which causes a slight reduction of sensitivity
where they are joined together. 

The PSRFITS scales and offsets of the Mock sub-bands are adjusted such that the
data value levels of top and bottom bands are appropriately weighted with
respect to each other.

The combining of the two Mock sub-bands is performed using \combinemocks of 
\psrfitsutils\footnote{https://github.com/scottransom/psrfits\_utils}.

Next, the sub-integrations containing the calibration diode signal are deleted
from the observation. The start time and length of the observation are updated
accordingly.

At this stage, prior to searching for periodic and impulsive signals, \presto's
\rfifind is run on the merged observation to generate an RFI mask. See
\sect~\ref{sec:masking} for details.

\subsection{Searching Components}
\label{sec:pipeline-search}

We will now cover the various steps required to search for pulsars and
transients. 

\subsubsection{Dedispersion}
\label{sec:ddplan}
Because the DMs of yet-undiscovered pulsars and transients are not known in
advance, a wide range of trial DMs must be used to maintain sensitivity to
pulsars. For each trial DM value a dedispersed time series is produced
    by shifting the frequency channels according to the assumed DM value and
then summing over frequency. When generating these time series, the motion of
the Earth around the Sun is removed so that the data are referenced to the
Solar System barycenter, assuming the coordinates of the beam center.

The PALFA \presto pipeline searches observations for periodic and impulsive
signals up to a DM of $\sim$\,10000\,\dmunit. We search to such
high DMs despite the maximum DM in our survey region predicted by the \nemodel
model being \app{1350}\,\dmunit \citep{cl02} to ensure sensitivity to highly-dispersed,
potentially extragalactic FRBs \citep[e.g.][]{tsb+13,sch+14}.

A dedispersion plan is determined by balancing
the various contributions to pulse broadening that can be controlled: the
duration of each sample (including down-sampling), $\tau_\mathrm{samp}$; the
dispersive smearing within a single channel, $\tau_\mathrm{chan}$; the
dispersive smearing within a single sub-band due to approximating the DM,
$\tau_\mathrm{sub}$; and the dispersive smearing across the entire observing
band due to the finite DM step size (i.e.  if the DM of the pulsar is half-way
between two DM trials), $\tau_\mathrm{BW}$. Additionally, pulses are broadened by
interstellar scattering, $\tau_\mathrm{scatt}$, which cannot be removed. 
The amount of scatter-broadening depends on the DM, observing frequency and
line-of-sight.  \citet{cor02} empirically determined the relationship
as

\begin{eqnarray}
    \label{eq:scattering}
    \log \tau_\mathrm{scatt} = &-&3.59 + 0.129 \log \mathrm{DM} \nonumber \\
                        &+& 1.02 \left( \log \mathrm{DM} \right)^2 - 4.4 \log \nu,
\end{eqnarray}

\noindent where $\tau_\mathrm{scatt}$ is given in $\mu$s, and $\nu$ is the
    observing frequency in GHz. Even for the same DM, $\log
\tau_\mathrm{scatt}$ are different for pulsars in different locations with a
standard deviation of $\sigma=0.65$ \citep{cor02}. Because
$\tau_\mathrm{scatt}$ cannot (in practice) be corrected, we ignore it when
determining our dedispersion plan.

    The total correctable pulse broadening, $\tau_\mathrm{tot}$, is estimated
    by summing the first four contributions in quadrature, 

\begin{equation}
    \tau_\mathrm{tot} = \sqrt{\tau_\mathrm{samp}^2 + \tau_\mathrm{chan}^2 +
                              \tau_\mathrm{sub}^2 + \tau_\mathrm{BW}^2}. 
    \label{eq:broadening}
\end{equation}

\noindent All of these broadening terms vary with DM. The dedispersion
    plan is chosen to equate these four broadening effects roughly by adjusting
    the DM step-size and down-sampling factor as a function of DM.  To reduce
    the number of DM trials, the step-size is never so small that
    $\tau_\mathrm{BW} \lapp 0.1$\,ms.

\begin{figure}[t!]
        \includegraphics[width=\columnwidth]{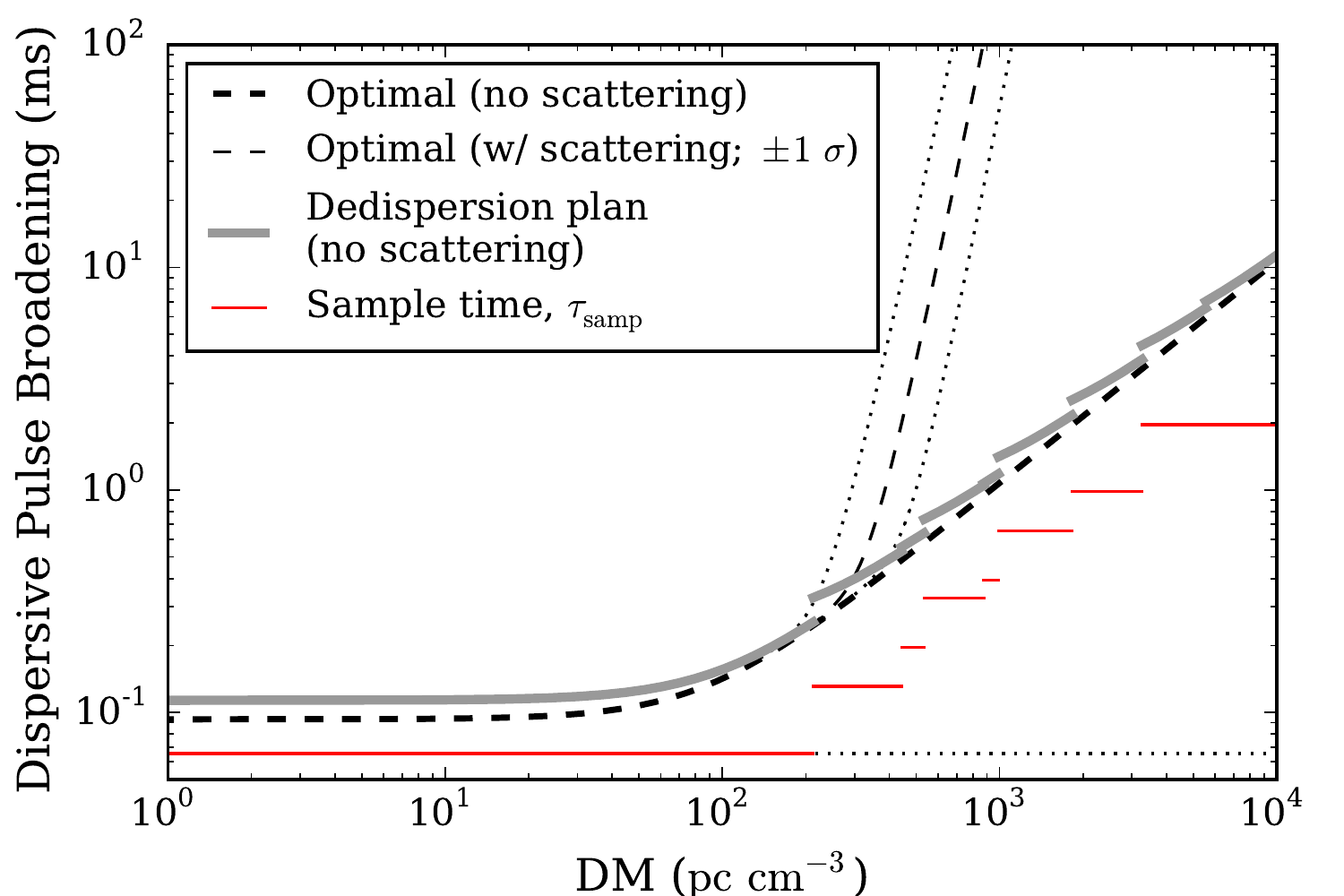}
        \caption{Pulse broadening from down-sampling,
                and dispersive DM smearing for the dedispersion plan generated
                by \ddplan shown in Table~\ref{tab:ddplan} (gray), as well as the
                optimal case (dashed black) where neither down-sampling nor
                smearing from DM errors are included. The optimal case
                including interstellar scattering is shown (with $\pm~1\sigma$;
                thin dashed black) assuming the empirical scattering
                dependence on DM of \citet{cor02}. While this dependence is
                likely reasonable for estimating the scattering of Galactic
                sources, it is likely to grossly overestimate the scattering of
                extragalactic sources (e.g.  FRBs). In all cases, the middle of
                the observing band is assumed (\app{1375}\,MHz).
                Discontinuities are due to down sampling. The horizontal lines
                (red) show the down sampled time resolution at various DMs.
    \label{fig:ddplan}}
\end{figure}

The PALFA survey dedispersion plan for Mock spectrometer data was determined
with a version of \presto's \ddplan modified to allow for non-power-of-two
down-sampling factors, and is shown in Table~\ref{tab:ddplan}. The
down-sampling factors are selected to be divisors of the number of spectra per
sub-integration, 15270.  The amount of dispersive smearing incurred at the
middle of the observing band, \app{1375}\,MHz, when using the dedispersion plan
in Table~\ref{tab:ddplan}, ranges from $\sim$\,0.1\,ms for the lowest DMs, to
$\sim$\,1\,ms for DMs of a few 100\,\dmunit, increasing to $\sim$\,10\,ms for a
DM of $\sim$\,10000\,\dmunit. Above a DM of $\sim$\,500\,\dmunit scattering
begins to dominate (see Fig.~\ref{fig:ddplan}). 

The more aggressive down-sampling at higher DMs has the advantage of reducing
the data size, making the analysis more efficient. Also, at higher DMs the
step-size between successive DM trials is increased, further reducing the
amount of processing.  Therefore, the extra computing required to go to high
DMs is relatively small compared to what is required to search for pulsars and
transients at low DMs. Searching DMs between 1000\,\dmunit and 10000\,\dmunit
adds only \app{5}\,\% the total data analysis time.

Dedispersion is done with \presto's \prepsub, passing through the raw data 99
times, and resulting in 7292 dedispersed time series. In all cases \prepsub
internally uses 96 sub-bands, each of 10\,MHz, for its two-stage sub-band
dedispersion process. Time intervals containing strong impulsive RFI are removed
by \prepsub, as prescribed by a RFI mask (see \sect~\ref{sec:masking}). 

A second set of dedispersed time series are created as before, but also
applying a version of the zero-DM filtering technique described by
\citet{ekl09} that has been augmented to use the bandpass shape when
removing the zero-DM signal from each channel. These zero-DM filtered time
series are especially useful for single-pulse searching, which is described in
\sect~\ref{sec:singlepulse}. See \sect~\ref{sec:zerodm} for details on
time-domain RFI mitigation strategies used. 

Dedispersion makes up roughly 15-20\,\% of the processing time.


\subsubsection{Periodicity Searching}
\label{sec:periodicity}

For every dedispersed time series, the discrete Fourier transform (DFT) is computed
using \presto's \realfft. Prior to searching the DFT for peaks, it is normalized to have
unit mean and variance. The normalization algorithm is designed mainly to suppress red noise (i.e. low-frequency trends in
the time series; for more details see \sect~\ref{sec:deredden}). Also, Fourier bins likely to
contain interference are replaced with the median-value of nearby bins. Details
of the algorithm used to determine RFI-prone frequencies are described in
\sect~\ref{sec:zapping}. 

Two separate searches of the DFT are conducted using \presto's \accelsearch.
Both searches identify peaks in the DFT down to a frequency of 0.125\,Hz.

The
first, \textit{zero-acceleration}, search is tuned to identify isolated
pulsars. The power spectrum of the signal from an isolated pulsar will consist
of narrow peaks at the rotational frequency of the pulsar and at harmonically
related frequencies. The number of significant harmonics depends on the width
of the pulse profile, $W$, and the spin period, $P$, as $N_\mathrm{harm} \sim
P/W$. To improve the significance of narrow signals, power from harmonics is
summed with that of the fundamental frequency. The zero-acceleration search
sums up to 16 harmonics, including the odd harmonics, in powers of 2 (i.e. 1,
2, 4, 8, 16 harmonics). This harmonic summing procedure also improves the
precision of the detected frequency.

The second, \textit{high-acceleration} search is optimized to find pulsars in
binary systems. The time-varying line-of-sight velocity of such pulsars gives
rise to a Doppler shift that varies over the course of an observation. This
smears the signal over multiple bins in the Fourier domain.  To recover
sensitivity to binary pulsars we use the Fourier-domain acceleration search
technique described in \citet{rem02}. In short, the high-acceleration search
performs matched-filtering on the DFT using a series of templates each
corresponding to a different constant acceleration.  We search using templates
up to 50 Fourier bins wide, which corresponds to a maximum acceleration of
$\sim$\,1650\,\mss for a 5-min observation of a 10-ms pulsar.  Only up to 8
harmonics are summed in the high-acceleration case because of its larger
computational requirements.

For each of the periodic signal candidates identified in both the zero- and
high-acceleration searches we compute the equivalent
Gaussian significance, \sigfourier, based on the probability of seeing a noise value with
the same amount of incoherently summed power \citep[see][for details]{rem02}.
The zero- and high-acceleration candidate information is saved to separate
lists for later post-processing (see \sect~\ref{sec:sifting}).

Typically, the zero-acceleration and high-acceleration searches make up between
2\,\%-5\,\% and \app{30}\,\% of the overall computation time, respectively. 

\subsubsection{Single Pulse Searching}
\label{sec:singlepulse}

Each dedispersed time series is also searched with \presto's \spsearch for
impulsive signals with a matched-filtering technique \citep[e.g.][]{cm03}.
Multiple box-car templates corresponding to a range of durations up to 0.1\,s
are used. Candidate single-pulse events at least 5~times brighter than the
standard deviation of nearby bins, \sigmaloc, are recorded. Diagnostic plots
featuring only $>$\,6\,\sigmaloc candidate events are generated and archived
for later viewing.  In addition to the basic diagnostic plots, all of the
$>$\,5\,\sigmaloc events are used in post-processing algorithms designed to
distinguish astrophysical signals (e.g.  from pulsars/RRATs and extragalactic
FRBs) from RFI and noise.  The algorithms employed by PALFA are described
elsewhere \citep{spi13,kkl+15}. 

The same searching and post-processing procedure is also applied to data from
which the DM=0\,\dmunit time series has been subtracted, using an enhanced
version of what was originally described in \citet{ekl09}. See
\sect~\ref{sec:zerodm} for more details about the time-domain RFI-mitigation
techniques used.

The single-pulse searching makes up approximately 20\,\% of the computing time.

\subsubsection{Sifting}
\label{sec:sifting}

As described above, the output of periodicity searching is a set of files, the
zero- and high-acceleration candidate lists for each DM trial, containing the
frequency of significant peaks found in the Fourier transformed time series,
along with other information about the candidate. 
In total, for all DMs, there are typically $\sim\,10^4$ period-DM pairs per beam. 
These signal candidates are \textit{sifted} to identify the most promising
pulsar candidates, match harmonically related signals, and reject RFI-like
signals. 

The first stage of the sifting process is to remove short-period candidate
signals ($P < 0.5$\,ms), which contribute a large number of false-positives, as
well as to ensure no candidate signals with periods longer than the limit of our search
($P > 15$\,s) are present.  Weak candidates with Fourier-domain significances $\sigfourier
< 6$ are also removed.  Furthermore, candidates with weak or strange harmonic
powers are rejected if they match one of the following cases: 1) the candidate
has no harmonics with $\sigfourier > 8$; 2) the candidate has \gapp\,8 harmonics
and is dominated by a high harmonic (fourth\footnote{We number harmonics such
that the frequency of the Nth harmonic is N times larger than the fundamental
frequency.} or higher), having at least twice as much power as the
next-strongest harmonic; 3) the candidate has 4 harmonics and is dominated by a
high harmonic (third or higher), having at least three times as much power as
the next-strongest harmonic.

The next stage of sifting is to group together candidates with similar periods
(at most 1.1 Fourier bins apart) found in different DM trials. When a duplicate
period is found, the less significant candidate is removed from the main list, and its DM is
appended to a list of DMs where the stronger candidate was detected. 

At this stage, for each periodic signal, there is a list of DMs at which it was
detected. The next step is to purge candidates with suspect DM detections.
Specifically, candidates not detected at multiple DMs, candidates that were most
strongly detected at DM~$\leq 2$\,\dmunit, and candidates that were not detected
in consecutive DM trials are all removed from subsequent consideration.

The steps described above are applied separately to candidates found in
the zero- and high-acceleration searches. At this point, the two candidate
lists are merged, and signals harmonically related to a stronger candidate are
removed from the list. This process checks for a conservative set of
integer harmonics, and small integer ratios between the signal
frequencies. As a result, some harmonically related signals are occasionally
retained in the final candidate list.

The sifting process typically results in \app{200} good candidates per beam, 
of which $\sim\!100$ are above the significance threshold for folding.
The fraction of time spent on candidate sifting is negligible
($<0.1$\,\%) compared to the rest of the pipeline.


\subsubsection{Folding}
\label{sec:folding}
The raw data are folded for each periodicity candidate with
$\sigfourier \geq 6$ remaining after the sifting procedure using \presto's
\prepfold. At most 200 candidates are folded for each beam. In more than 99\,\%
of cases this limit is sufficient to fold all $\sigfourier \geq 6$ candidates.
If too many candidates have $\sigfourier \geq 6$, the candidates with largest
\sigfourier are folded.

After folding, \prepfold performs a limited search over period,
period-derivative, and DM to maximize the significance of the candidate.
However, for candidates with $P>50$\,ms the search over DM is excluded because
it is prone to selecting a strong RFI signal at low DM even if there is a
pulsar signal present. Furthermore, the optimization of the period-derivative
is also excluded for $P>500$\,ms candidates.

For each folded candidate a diagnostic plot is generated \citep[see][for
examples]{ran01}. These plots, along with basic information about the candidate
(optimized parameters, significance, etc.) are placed in the PALFA processing
results database, hosted at the Cornell Center for Advanced Computing. The
\prepfold binary output files generated for each fold are also archived at
Cornell.

The binary output files created by \prepfold are used by a candidate-ranking
artificial intelligence system, as well as to calculate heuristics for
candidate sorting algorithms. Details can be found in \sect \ref{sec:pipeline-post}.

Folding the raw data for up to 200 candidates per beam is a
considerable fraction (\app{25}\,\%) of the overall computing time.

\subsection{RFI-Mitigation Components}
\label{sec:pipeline-rfi}

The sensitivity of Arecibo and PALFA can only be fully realized if
interference signals in the data are identified and removed. To work toward
this goal, the PALFA pipeline includes multiple levels of RFI excision. Each
algorithm is designed to detect and mitigate a different type of terrestrial
signal. Because these interference signals are terrestrial they are not
expected to show the $1/f^2$ frequency sweep characteristic of interstellar
signals. Unfortunately, some broadband terrestrial signals show
frequency sweeps that cannot be distinguished from astronomical signals by
data analysis pipelines \citep[e.g. ``perytons''][]{bbe+11}. Despite some
non-astronomical signals remaining in the data, the suite of RFI-mitigation
techniques described here are an essential part of the pipeline. 

All of the algorithms described here are applied to non-dedispersed,
topocentric data.

\subsubsection{Removal of Site-Specific RFI}
\label{sec:radar}
\begin{figure}[t!]
        \includegraphics[width=\columnwidth]{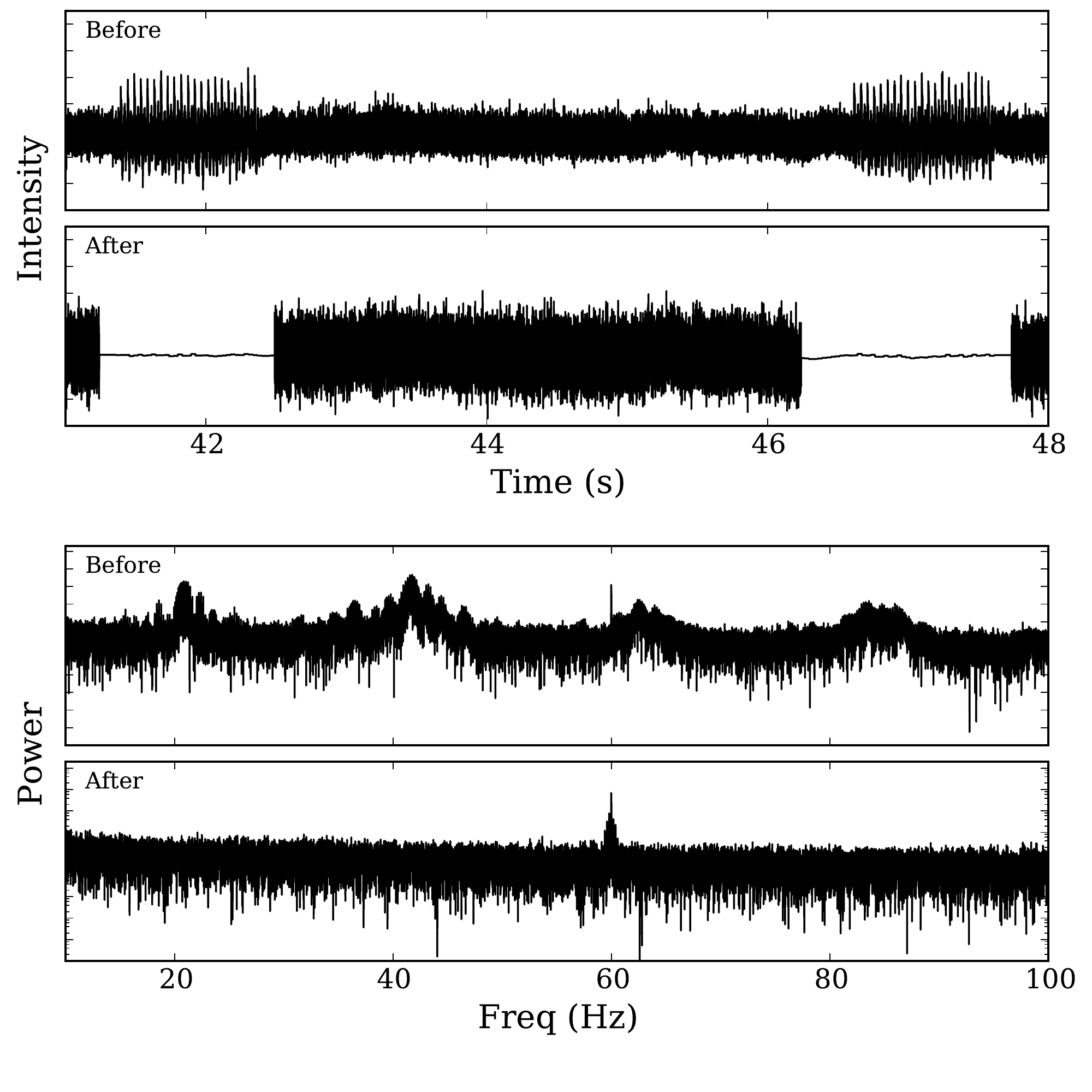}
        \caption{
            An example of the effect of the bursts of interference
                caused by some of the electronics equipment at the Arecibo
                Observatory on PALFA survey data in time and frequency domain
                (labeled ``Before'') and the same interval of time series and
                power spectrum after our finely-tuned removal algorithm,
                described in \sect~\ref{sec:radar}, is applied (labeled
                ``After''). Part of the time series is sacrificed, but the
                broad features in the frequency domain are completely removed.
                The RFI peak at 60\,Hz that remains in the bottom panel is
                caused by the electrical mains and is later removed by zapping
                intervals of the power spectrum (described in
        \sect~\ref{sec:zapping}). The source of this interference signal has been
        identified and can be dealt with by shutting it off during PALFA observations.
        \label{fig:radar}}
\end{figure}

Unfortunately, some of the electronics hardware at the Arecibo Observatory,
specifically the ALFA bias monitoring system, introduced strong periodic
interference into our data. By the time the source of the interference was
determined several months of observations had been affected. Fortunately, we
were able to develop a finely-tuned algorithm to excise the signal using our
knowledge of the sub-pulse structure to identify and remove these intense
bursts of interference. Finely tuned algorithms such as this one have the
advantage of more easily identifying specific RFI signals and only extracting
the affected data. In this particular case, each 1-s burst of RFI is made up of
a comb of
\app{10}\,ms-long sub-pulses with repeating every \app{50}\,ms. By
removing these bursts, our algorithm largely eliminates the broad peaks in the
Fourier domain that are introduced by the pernicious electronics, typically
between 1 and 1000~Hz (i.e. exactly where we expect pulsars to be found). See
Figure~\ref{fig:radar} for an example.  Furthermore, by removing the
interference pulses in the time domain, the power spectrum is cleaned without
sacrificing any intervals of the Fourier domain, as would be the case with the
zapping algorithm described in \sect~\ref{sec:zapping}.

Because the equipment causing the bursts of interference in our
observations is not essential to data taking we have been able to shut it
off during PALFA sessions.

\subsubsection{Narrow-Band Masking}
\label{sec:masking}
Every observation is examined for narrow-band RFI signals using
\presto's \rfifind, which considers 2-s long blocks of data in each frequency
channel separately.
For each block of data two time-domain
statistics are computed: the mean of the block data value, and the standard
deviation of the block data values.  Also, one Fourier-domain statistic is
computed for each block: the maximum value in the power spectrum.  Blocks where
the value of one or more of these three statistics is sufficiently far from the
mean of its respective distribution are flagged as containing RFI. For the two
time-domain metrics, in the PALFA survey the threshold for flagging a block is
10 standard deviations from the mean of the distribution, and for the
Fourier-domain metric, the threshold is 4 standard deviations from the mean.
The resulting list of flagged blocks is used to mask out RFI.  Masked blocks
are filled with constant data values chosen to match the median bandpass.
Channels that are more than 30\,\% masked are completely replaced, as are
sub-integrations that are at least 70\,\% masked.

On average, only \app{5.75}\,\% of time-frequency space is masked by this
algorithm, and \app{93}\,\% of observations have mask-fractions less than 10\,\%.
Observations where the mask-fraction is larger than 15\,\% will be re-inserted into
the list of sky positions to observe. These represent only \app{1.1}\,\% of observations.

The fraction of data masked for each beam, and a graphical representation of
the mask are stored in the results database as diagnostics of the observation
quality. 

Generating the \rfifind mask makes up only \app{1}\,\% of the total
pipeline running time.

\subsubsection{Time-Domain Clipping and Filtering}
\label{sec:zerodm}

It is possible for broad-band impulsive interference signals to be missed by
the masking procedure described above if the signals are not sufficiently
strong to be detected in individual channels. Fortunately, the PALFA pipeline
makes use of a complementary algorithm designed to remove such signals from the
data: a list of bad time intervals is determined by identifying samples in the
DM=0\,\dmunit time series that are significantly larger ($>6\,\sigmaloc$) than the
surrounding data samples. The spectra corresponding to the bad time intervals
are replaced by the local median bandpass.

\begin{figure}[t!]
        \includegraphics[width=\columnwidth]{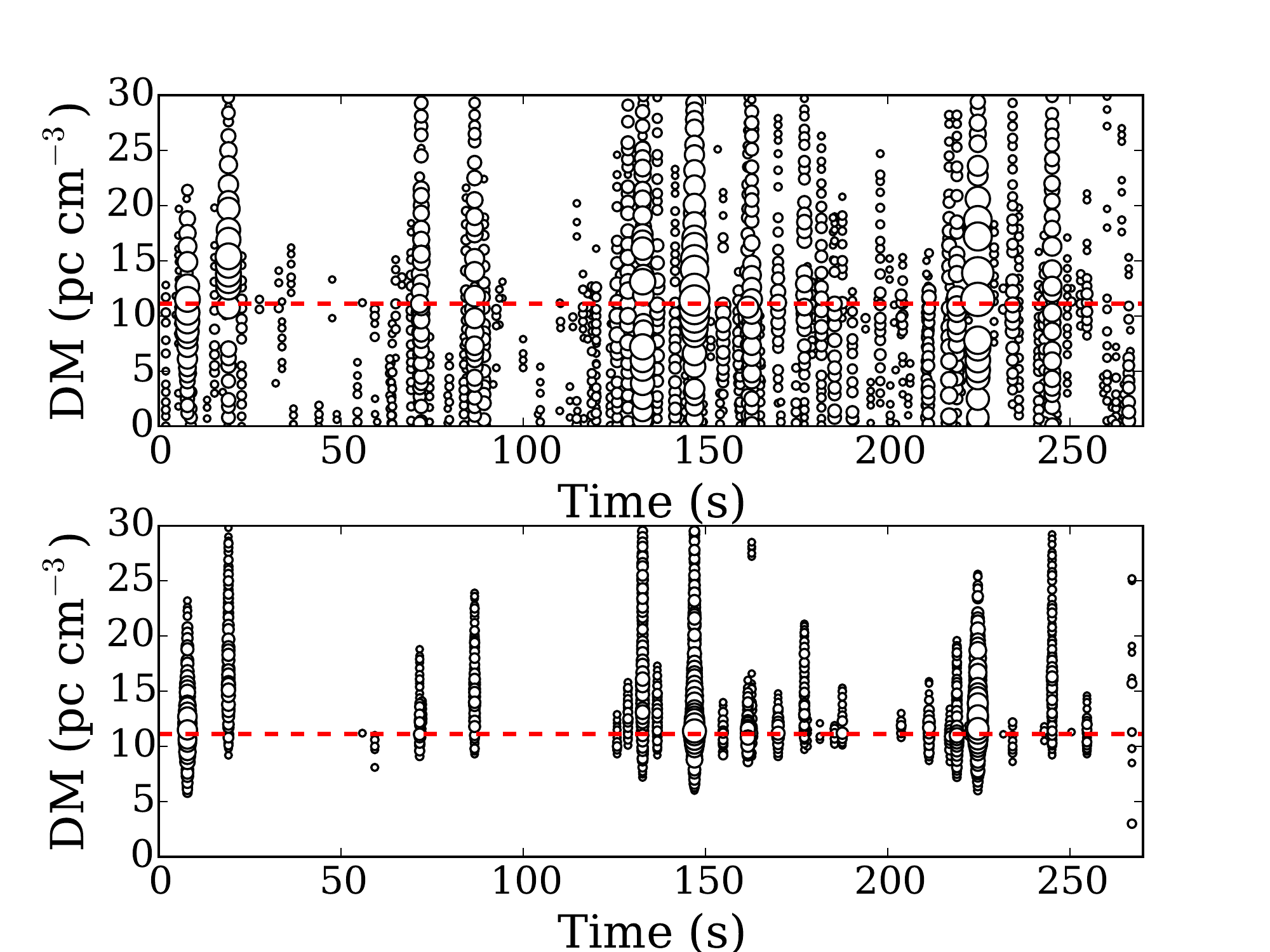}
        \caption{Comparison of single-pulse events detected in a PALFA
        observation of PSR~J1908+0734 in a search of the un-filtered time
        series (top) and the zero-DM filtered time series (bottom).
        Each circle represents the time and DM of an impulsive signal
        found by \presto's \spsearch. The size of the circle is proportional to
        the significance of the signal (up to a maximum radius).  Most of the
        RFI is filtered out of the observation by the zero-DM algorithm while
        leaving the pulsar pulses, albeit with some loss of significance at the
        lower DMs.  Thus, the zero-DM filtering technique makes it far easier
        to disentangle astrophysical signal at non-zero DMs from RFI at
        DM=0\,\dmunit both by eye and algorithmically. The pulsar's
        DM=11\,\dmunit is indicated with the dashed red line.
        \label{fig:zerodm_comparison}}
\end{figure}

As previously mentioned, for single-pulse searching, the PALFA pipeline also
applies the \presto-implementation of the zero-DM filtering technique described
in \citet{ekl09}. This implementation enhances the original prescription by
using the bandpass shape as weights when removing the DM=0\,\dmunit signal.
The zero-DM filter greatly reduces the impact of RFI on single-pulse searching,
facilitating low-DM RRATs being distinguished from RFI. To illustrate the
benefits of zero-DM filtering, Figure~\ref{fig:zerodm_comparison} shows a
comparison of the single-pulse events identified by \spsearch in an observation
of PSR~J1908+0734 with and without filtering.

\subsubsection{Red-Noise Suppression}
\label{sec:deredden}
In order to properly normalize the power spectrum and compute more
correct false-alarm probabilities \citep[see][]{rem02}, we use a power spectrum
whitening technique to suppress frequency-dependent, and in particular ``red''
noise.  The median power level is measured in blocks of Fourier frequency bins
and then multiplied by $\log 2$ to convert the median level to an equivalent
mean level assuming that the powers are distributed exponentially (i.e.
$\chi^2$ with 2 degrees-of-freedom).

The number of Fourier frequency bins per block is determined by the
log of the starting Fourier frequency bin, beginning with 6 bins and
increasing to approximately 40 bins by a frequency of 6\,Hz.  Above
that frequency, where there is little to no ``colored'' noise, block
sizes of 100 bins are used.  The resulting filtered power spectrum has
unit mean and variance.  This process is accomplished with \presto's
\rednoise program.

\subsubsection{Fourier-Domain Zapping}
\label{sec:zapping}
Sufficiently bright periodic sources of RFI can be mistakenly identified as
pulsar candidates by our FFT search. To excise, or \textit{zap}, these signals
from our data we tabulate frequency ranges often contaminated by RFI. The
Fourier bins contained in this \textit{zap list} are replaced by the average of
nearby bins prior to searching. 

\begin{figure}[t!]
        \includegraphics[width=\columnwidth]{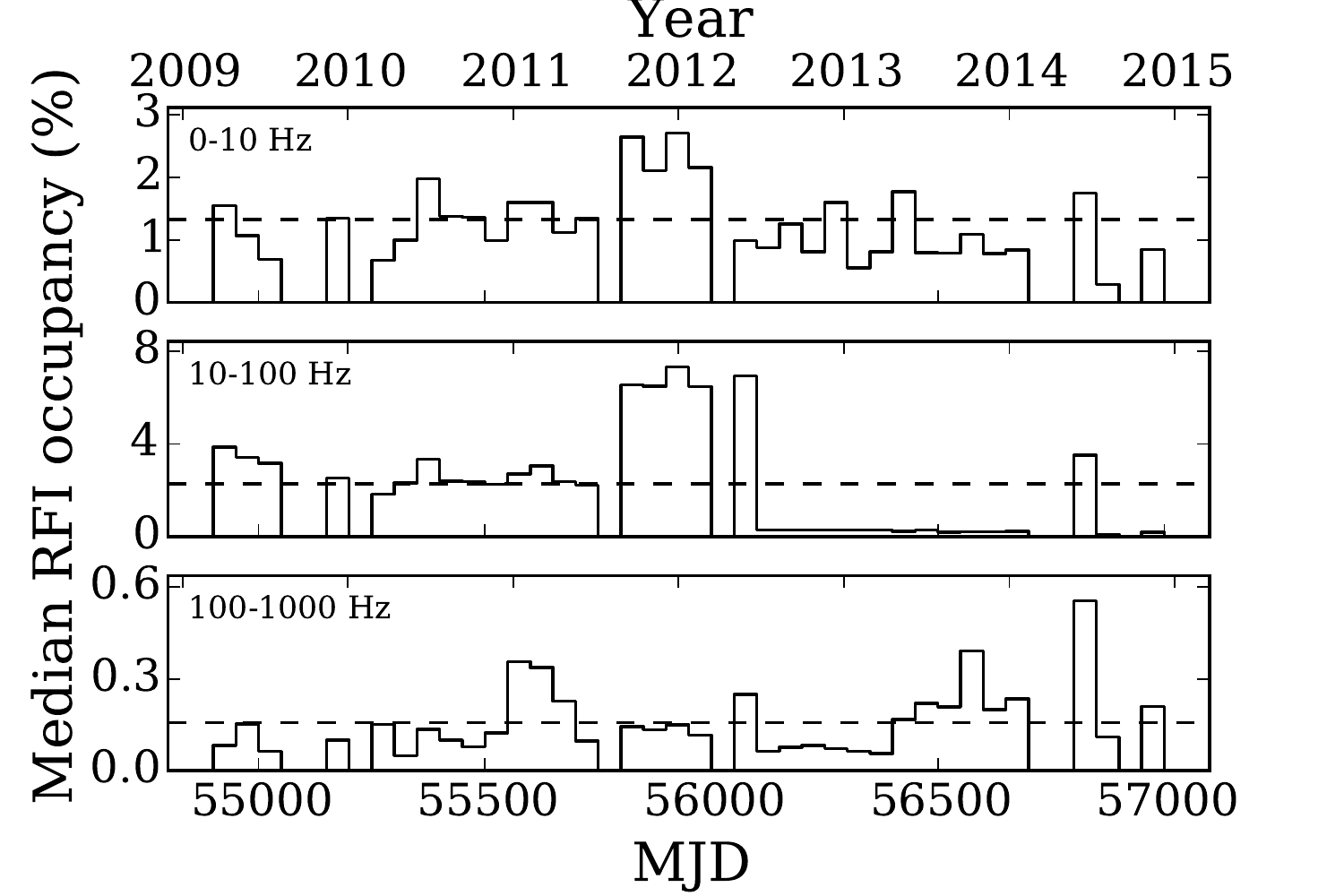}
        \caption{ 
                Median percentage of the Fourier domain occupied by RFI in
                three frequency ranges for 50-day intervals (solid lines)
                compared against the median percentage for all observations
                (dashed lines). Many periodic sources of RFI are found to vary
                on daily time scales.  Thus, lists of RFI-contaminated
                    Fourier frequencies to be removed from the power spectrum
                    prior to searching are tailored to the RFI of each MJD.
                    The increase in RFI in the middle panel between
                        MJD 55750 and 56100 was due to on-site electronics at
                the telescope, which since being identified in 2012 June (MJD
        $\simeq 56100$) has nearly always been turned off during PALFA
observations, significantly reducing the RFI in the 10-100\,Hz interval.
        \label{fig:zapfrac_summary}}
\end{figure}

The RFI environment at Arecibo is variable. The number, location, and width of
interference peaks in the Fourier transform of DM=0~\dmunit time series vary on a
time scale of months to years. To demonstrate this, the fraction of Fourier bins
occupied by RFI as a function of epoch is illustrated in
Figure~\ref{fig:zapfrac_summary}.  The median fraction of the Fourier spectrum
occupied by RFI for all Mock spectrometer data for various intervals is: 2.9\,\%
(0-10\,Hz), 5.1\,\% (10-100\,Hz), and 0.5\,\% (100-1000\,Hz).  To account for this
dynamic nature of the RFI, we compute zap lists for each MJD. 

To compute zap lists we exploit the fact that RFI signals are typically
detected by multiple feeds in a single 5-min pointing, or persist for most of
an observing session (typically 1--3 hours). The strategy we employ
here is similar to what was used in the Parkes Multibeam Pulsar Survey
\citep{mlc+01}. Fourier bins contaminated by RFI are determined by finding
peaks in a \textit{median power spectrum}, which is comprised of the bin-wise
median of multiple DM=0\,\dmunit power spectra. This is done twice, using two
different subsets of data: a) all observations made with a given ALFA feed on a
given day (to identify RFI signals that persist for multiple hours, or issues
specific to the ALFA receiver), and b) all seven observations from a given
pointing (to identify shorter-duration periodic RFI signals that enter multiple
feeds). The zap list for any given observation is the union of the lists for
its pointing and its feed. 

With the advent of sophisticated candidate ranking and candidate classifying
machine-learning algorithms (see \sect~\ref{sec:pipeline-post}), it is better to
leave some RFI in the data than to remove large swaths of the Fourier domain.
To avoid excessive zapping we remove at most 3\,\% from each frequency decade,
up to a maximum of 1\,\% globally, preferentially zapping bins containing the
brightest RFI.

In addition to being an essential part of the PALFA RFI-mitigation strategy,
zap lists have also proven to be a useful diagnostic for monitoring the
RFI environment at Arecibo.

\subsection{Post-processing Components}
\label{sec:pipeline-post}

\subsubsection{Ratings}
\label{sec:ratings}

A series of 19 
heuristic ratings are computed for each folded periodicity
candidate produced by the data analysis pipeline. These ratings encapsulate
information about the shape of the profile, the persistence and broadbandedness
of the signal, whether the frequency of the signal is particularly RFI-prone,
and whether the signal is stronger at DM=0\,\dmunit. Each of the ratings is
uploaded to the results database, and is available for querying and sorting
candidates (see \sect~\ref{sec:cyberska}). The ratings and brief descriptions
are presented in Table~\ref{tab:ratings}.

The ratings are incorporated into candidate-selection queries along with
standard parameters such as period, DM, and various measures of time-domain and
frequency-domain significance. Using ratings in this way allows users to
constrain the candidates they view to have certain features they would require
when selecting promising candidates by eye. Alternatively, the ratings have been
used in a decision-tree-based artificial intelligence (AI) algorithm, but this has
since been supplanted by the more sophisticated ``PICS'' algorithm described
in \sect~\ref{sec:ai} \citep{zbm+14}. 

The code to compute the ratings\footnote{Available at
https://github.com/plazar/ratings2.0.} is compatible with the binary files
produced by \presto's \prepfold for each periodicity candidate. For each
candidate a text file is written containing the name, version, description, and
value for all ratings being computed. This task is performed as part of the
data analysis pipeline.  The rating information is later uploaded
to the results database. In cases where a new rating is devised,
or an existing rating is improved, the \prepfold binary files are fetched from
the results archive, ratings are computed in a stand-alone process (i.e.
independent of the pipeline), and the values are inserted into the database.
The values of improved ratings are inserted alongside values from old versions
to permit detailed comparisons.

\subsubsection{Machine Learning Candidate Selection}
\label{sec:ai}

All periodicity candidates are also assessed by the Pulsar Image-based
Classification System \citep[PICS;][]{zbm+14}, an
image-pattern-recognition-based machine-learning system for selecting
pulsar-like candidates. The PICS deep neural network enables it to recognize and
learn patterns directly from 2-D diagnostic images produced for every
periodicity pulsar candidate.  The large variety of pulsar candidates used to
train PICS has developed its ability to recognize both pulsars and their
harmonics.

PICS can reduce the number of candidates to be inspected by human experts by a
factor of \app{100} while still identifying 100\,\% of pulsars and 94\,\% of
harmonics to the top 1\,\% of all candidates \citep{zbm+14}. 

Since late 2013, PICS has been integrated directly into the PALFA processing
pipeline. It produces a single rating for each candidate, which is uploaded into the
results database as a rating (see \sect~\ref{sec:ratings}). So far, this has
aided in the discovery of 9 pulsars (see \sect~\ref{sec:results}).

\subsubsection{Coincidence Matching}
\label{sec:coincidence}

While PALFA has been successful at finding moderately bright MSPs, the vast
quantity of periodicity candidates close to the detection threshold at very
short periods (\lapp\,2\,ms) have made it more challenging to identify the
\textit{faint} MSPs in the PALFA results
database. To facilitate the process, a search for signals with compatible
periods, DMs and sky positions has been performed on the periodicity candidates
in the database. By applying our coincidence matching algorithm to the
complete list of folded candidates we are able to reliably probe lower
\snrs than would be reasonable to do thoroughly by manual viewing. This
algorithm is complementary to our machine learning technique that operates on
each candidate individually. The software developed to find matching
candidates is available on the web for general
use\footnote{https://github.com/smearedink/PALFA-coincidences}.

Large parts of the survey region have either been observed more than once or have
been densely sampled (see Fig.~\ref{fig:skymap}), making it possible to
match the detection of a pulsar from multiple observations confidently. For
each observation, a list of beams from other pointings that fall within $5'$
is generated.  Candidates from the different beams are matched by their DMs and
barycentric periods. Allowances are made for slightly different DMs and
periods, as well as for harmonically related periods.
Multiple matches that include the same candidate
are consolidated to form groups of more than two candidates.

The results of this matching algorithm are examined with a dedicated, web-based
interface.  Many known pulsars, especially high harmonics of very bright slow
pulsars, have already been identified.

As of 2015 January, 
our coincidence matching search has not yet resulted in the discovery of new
pulsars, but it continues to be applied to the results database. This algorithm
will be increasingly useful as more of the PALFA survey region becomes densely
sampled, and as more Mock spectrometer observations cover positions previously
observed with the WAPP spectrometers.

\subsection{Collaborative Tools}
\label{sec:cyberska}

The PALFA Consortium has created and made use of several online collaborative
tools on the CyberSKA portal\footnote{http://www.cyberska.org} \citep{kab+11},
a website developed to help astronomers build tools and strategies for
large-scale projects in the lead-up to the Square Kilometre Array (SKA). 

The CyberSKA portal allows for third-party applications to be accessed
directly without a need for separate user authentication. Within this framework
several PALFA-specific applications were developed:

\textit{Candidate Viewer} --
The primary method for viewing and classifying PALFA candidates is by using the
CyberSKA \textit{Candidate Viewer} application. It allows users to access the
Cornell-hosted results database using form-based, free-text, and saved queries.
Queries include basic observation and candidate information (e.g. sky position,
period, DM, significance), as well as ratings (\sect~\ref{sec:ratings}), and the
PICS classifications (\sect~\ref{sec:ai}).  Users are presented with a series of
\prepfold diagnostic plots in sequence, one for each candidate matching the
query. By inspecting the plots, as well as other relevant information provided,
such as a histogram showing the number of occurrences of signals in the relevant
frequency range as well as a summary plot showing all the beam's periodic
signal candidates in a period-DM plot, the user can quickly classify
candidates. Classifications are saved to the database and can be easily
retrieved.

\textit{Top Candidates} -- 
Especially promising candidates found with the Candidate Viewer can be added to
the \textit{Top Candidates} application, which is designed to store the most
likely pulsar candidates. The application also allows collaboration members to
view and vote on which candidates should be subject to confirmation
observations, as well as help organize and track these observations and their
outcomes.

\textit{Survey Diagnostics} --
Optimizing the use of telescope time and computing resources is extremely
important for large-scale pulsar surveys such as PALFA. The \textit{Survey
Diagnostics} application automatically compiles a set of information and a set
of plots from various sources to help the project run smoothly. This includes
the status of data acquisition and reduction, the severity of the RFI
environment, and the quality of the data. 

\section{Results}
\label{sec:results}

The PALFA Survey has discovered 145 pulsars, including 19 MSPs and 11 RRATs,
and one FRB, as of 2015 March.  The \presto-based pipeline described in
\sect~\ref{sec:pipeline} has discovered 41 pulsars from their periodic emission,
5 RRATs from their impulsive emission, and re-detected another 60 pulsars that
were previously discovered with other PALFA data analysis pipelines. The other
pulsars found in the PALFA survey were discovered with the different data
analysis pipelines, such as the E@H and Quicklook pipelines
\citep{akc+13,sto13} which use complementary
RFI-excision and search algorithms, with dedicated transient searches, or in
earlier observations with the WAPP spectrometers using an earlier version of
the pipeline described here. Not all sky positions observed with the WAPP
spectrometers have been covered with the Mock spectrometers yet.

\begin{figure*}[t]
        \includegraphics[width=\textwidth]{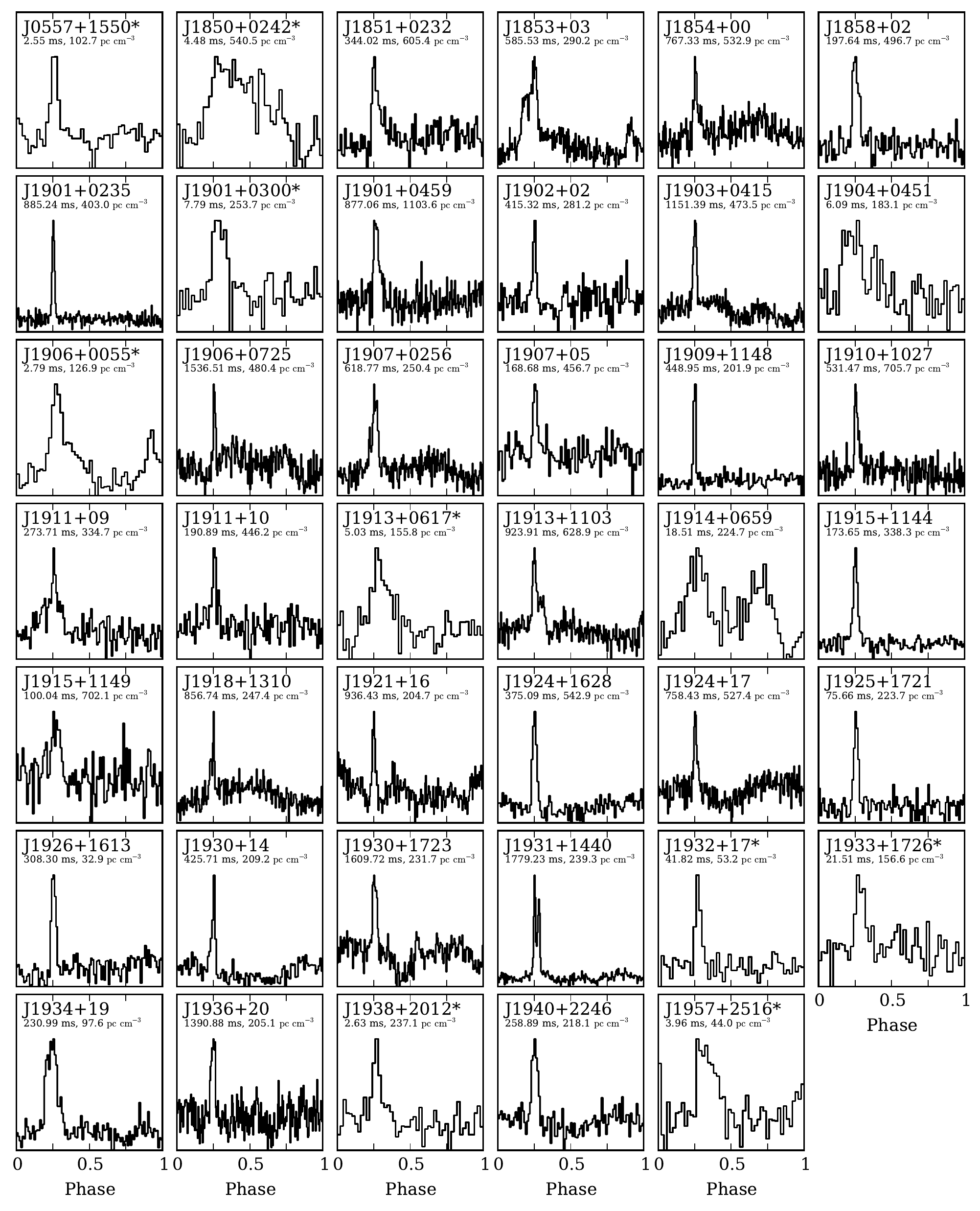}
        \caption{Pulse profiles at 1.4\,GHz from the discovery observations of
            the 41 pulsars discovered with the \presto-based PALFA pipeline in
            Mock spectrometer data. The name of each pulsar is included above
            each profile along with the period, and dispersion measure. The
            names of binary pulsars are indicated with an asterisk (*).  The number
            of bins across the profile is what was used by the pipeline, and is
            larger for longer period pulsars. These profiles also include
            intra-channel DM smearing, which is most significant for
            high-DM, short-period pulsars. The baselines of several profiles,
            predominantly of the long-period pulsars, show broad features due
            to interference and red noise in the data (for example, PSRs
            J1854+00, J1921+16, and J1930+1723).  The discovery profiles
            contaminated with RFI and red noise are shown here to highlight the
            ability of the PALFA pipeline to identify pulsars despite these
            conditions. Pulsars with truncated names do not yet have positions
            determined from timing campaigns.  
        \label{fig:discoveries}}
\end{figure*}

We report details for 41 of the periodicity-discovered
pulsars found in Mock spectrometer data with the pipeline described above. All
but one of these discoveries are in the inner Galaxy region.  These
pulsars were discovered by analyzing \nts{85333} beams, covering a total of
134\,\sqdeg, which consists of 80\,\sqdeg in the inner Galaxy region, and
54~\sqdeg in the outer Galaxy region (see Table~\ref{tab:data}).  Basic
parameters of the discoveries are in Table~\ref{tab:discoveries}, and pulse
profiles from the discovery observations are shown in
Figure~\ref{fig:discoveries}.

Eight of the 41 pulsars reported here are MSPs, 
including the most distant MSP (based on its DM) discovered to date,
PSR~J1850+0242. The distance estimated from the DM of PSR~J1850+0242, assuming
the NE2001 model \citep{cl02}, is 10.4\,kpc, a testament to the ability of the
PALFA survey to find highly dispersed, short period pulsars.
PSR~J1850+0242, along with three of the other MSPs discoveries reported
    here are described in detail in \citet{skl+15}. Three more of the MSPs
    reported here will be included in \nts{Stovall et al. (in prep.)}.

Nine of the 41 pulsars reported here are in binary systems, including
seven of the MSPs, and two slower pulsars, PSRs~J1932+17 ($P\simeq42$\,ms) and
J1933+1726 ($P\simeq22$\,ms), that were spun-up by the accretion of
mass and transfer of angular momentum, the so-called ``recycling'' process
\citep{acrs82}. The timing analysis of PSR~J1933+1726 will be provided by
\nts{Stovall et al. (in prep.)}

Timing solutions for six of the slow pulsars presented in this work,
including the young PSR~J1925+1721, will be published in a forthcoming paper
along with the timing of other PALFA-discovered pulsars (\nts{Lyne et al.,
in prep.}).


In addition to the 41 periodicity pulsars detailed here, the \presto-based
pipeline has found 5 RRATs. The beams containing these RRATs were identified
using a post-processing algorithm originally developed for pulsar surveys at
350\,MHz with the Green Bank Telescope \citep[see][for details]{kkl+15}.
Discovery parameters and detailed follow-up observations for these RRATs will
be described elsewhere.

\subsection{Estimating Flux Densities of New Discoveries}
\label{sec:faint}

The flux densities of the new discoveries were estimated using the radiometer equation
\citep{dtws85},

\begin{equation}
    S_\mathrm{est} = \frac{\snrtime \left ( T_\mathrm{sys} + T_\mathrm{sky} \right )}
                   {G(\theta, ZA) \sqrt{n_p t_\mathrm{obs} \Delta f}} 
              \sqrt{\frac{W}{P-W}},
    \label{eq:radiometer}
\end{equation}

\noindent where relevant parameters are the pulse profile width, $W$, the
telescope gain, $G(\theta, ZA)$, the number of polarization channels summed, $n_p$,
the observation length, $t_\mathrm{obs}$, the observing bandwidth, $\Delta f$,
the period of the pulsar, $P$, the system and sky temperatures,
$T_\mathrm{sys}$ and $T_\mathrm{sky}$, respectively.  
The time-domain signal-to-noise ratio, \snrtime, was measured from
folded profiles using the area under the pulse and the off-pulse RMS.

In some cases, predominantly for long-period pulsars, the baseline of the pulse
profile exhibited broad features, likely due to red noise. (See some examples
in Fig.~\ref{fig:discoveries}.) To more robustly estimate flux densities, we
fit Gaussian components to the pulse profile, including the broad off-pulse features.
The integrated pulsar signal was determined from the on-pulse components, and the noise
level of the profile was determined from the standard deviation of the residuals after
subtracting all fitted components from the profile.

The gain was scaled according to the angular offset of the pulsar from the beam
center, $\theta$, assuming an Airy disk beam pattern with $\mathrm{FWHM} =
3\farcm35$ \citep{cfl+06}, as well as the dependence on the zenith angle, $ZA$.
The gain also took into account the ALFA beam with which the pulsar was
detected, $G(0)=10.4$\,K/Jy for the central beam, and $G(0)=8.2$\,K/Jy for the
outer 6 beams \citep{cfl+06}.

Sky temperatures were scaled from the \citet{hssw82} 408-MHz survey to
1400\,MHz using a spectral index of $-2.76$ for the Galactic synchrotron
emission \citep{pbb+98}. The sky temperatures also include the 2.73\,K cosmic
microwave background.

\begin{figure}[t]
        \includegraphics[width=\columnwidth]{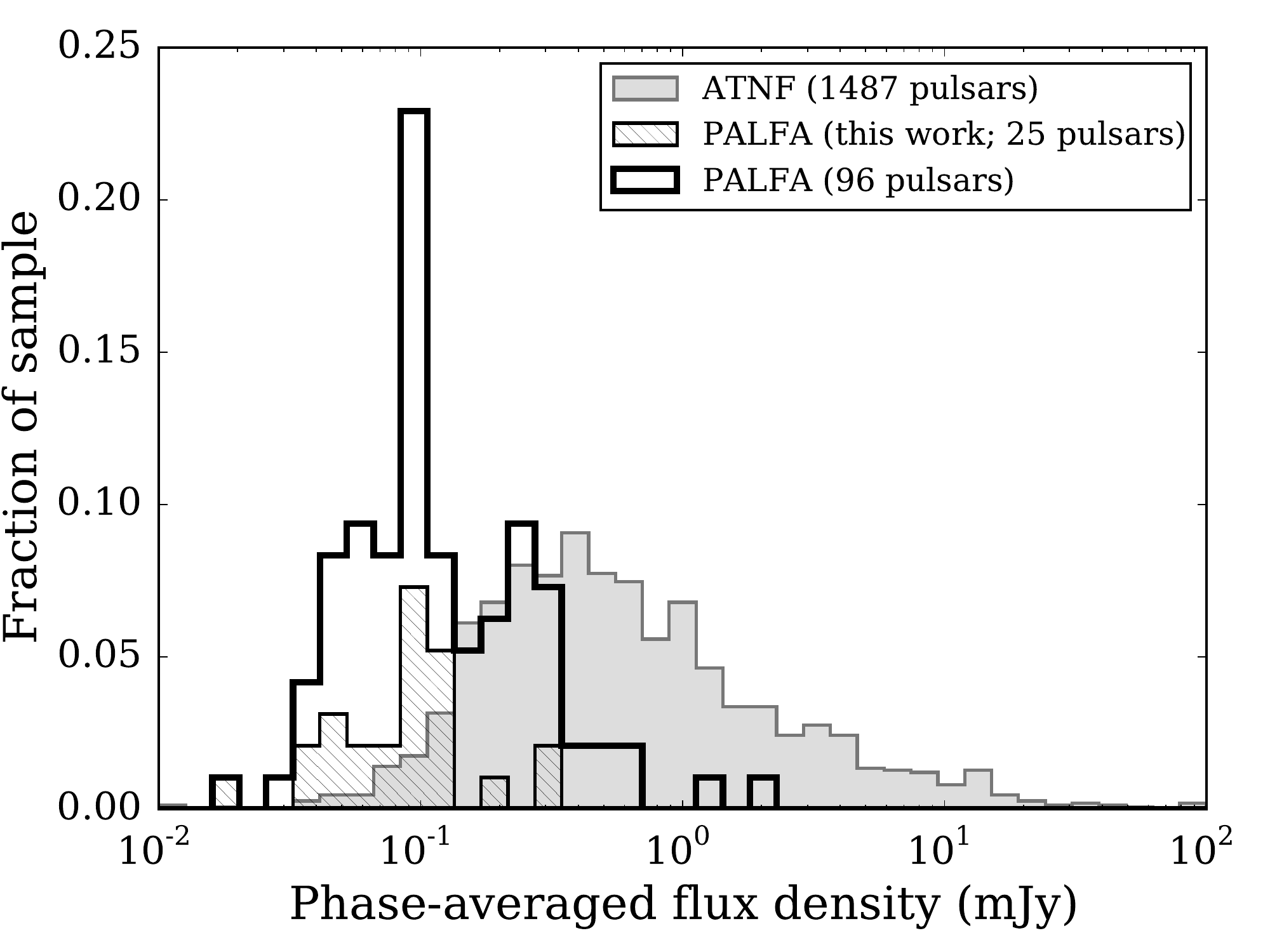}
        \caption{Distribution of phase-averaged flux densities of pulsars
        discovered in the PALFA survey, and the distribution of 1400-MHz
        phase-averaged flux densities from the ATNF pulsar catalogue of all
        non-PALFA, non-globular cluster discoveries. The sub-set of PALFA
        pulsars featured in this work is highlighted. Only PALFA-discovered
        pulsars with timing positions are included. 
        \label{fig:smean_hist}}
\end{figure}

The resulting phase-averaged flux density estimates of the PALFA pulsars
discovered with our pipeline range from 16\,$\mu$Jy to
280\,$\mu$Jy (see Table~\ref{tab:discoveries}), making them among the weakest
detected pulsars in the Galactic field, along with other PALFA-discovered
pulsars (see Fig.~\ref{fig:smean_hist}).

\subsection{Re-Detections of Known Pulsars}
\label{sec:known}

In total, 83 pulsars for which 1400-MHz phase-averaged flux densities, $S_{1400}$, are
reported in the ATNF catalogue were detected with the Mock spectrometers in 268
different PALFA observations (i.e. some known pulsars were re-detected multiple times). 

To confirm that our observing set-up is as sensitive as expected, we estimate the \snrtime
at which our pipeline should blindly re-detect known pulsars in our observations
and compare with the \snrtime measured from the profile of the corresponding
candidate. The expected \snrtime values were estimated by inverting
Eq.~\ref{eq:radiometer} to solve for the signal-to-noise ratio using $S_{1400}$
from the ATNF catalogue. As in \sect~\ref{sec:faint} the telescope gain is
modeled as an Airy disk with $\mathrm{FWHM}=3\farcm35$. 

\begin{figure}[t]
        \includegraphics[width=\columnwidth]{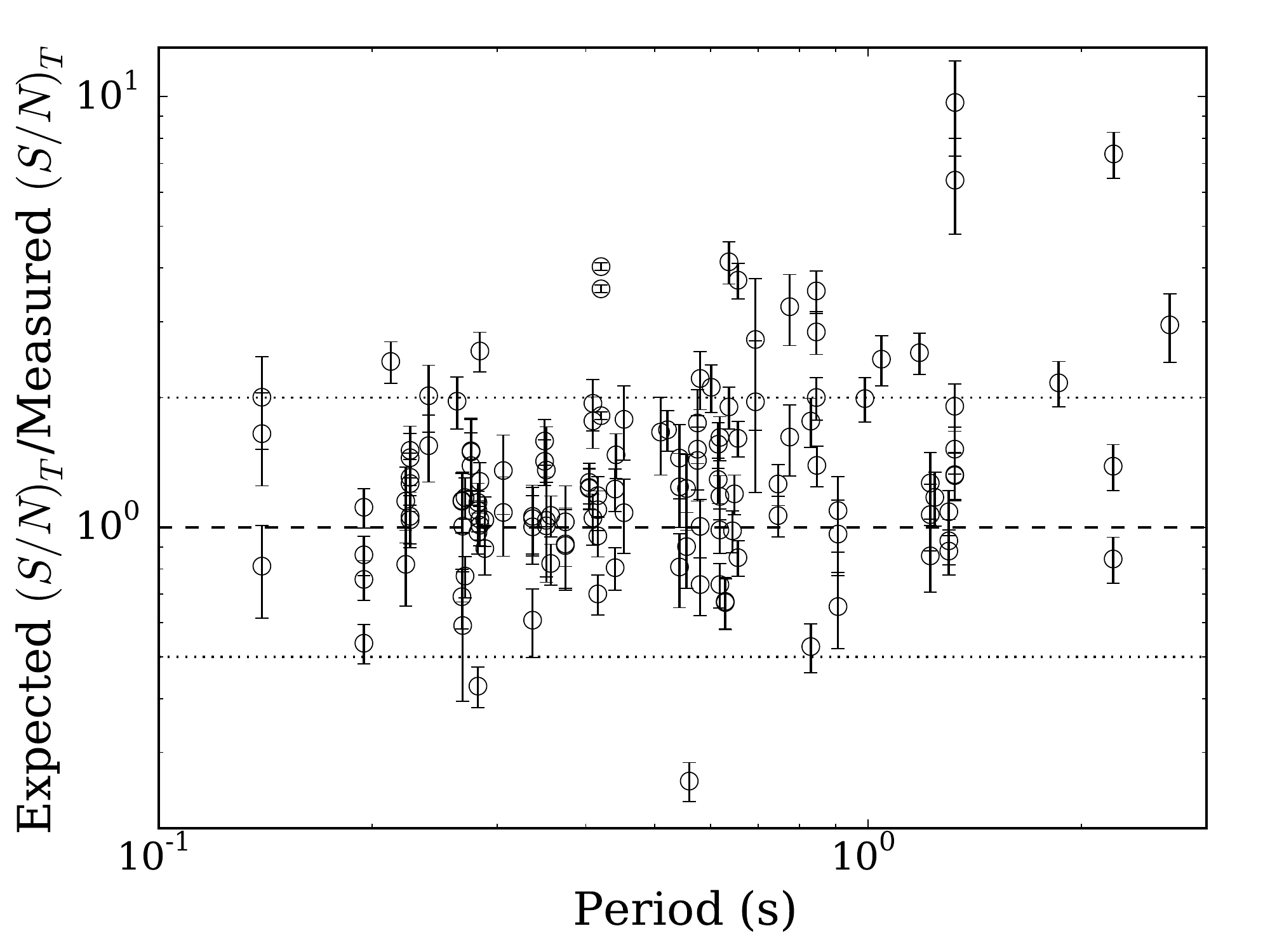}
        \caption{
            Ratio of expected and measured \snrtime as a function of pulsar period.
            Expected \snrtime values are calculated using the radiometer
            equation and measured flux densities at 1400\,MHz from the ATNF
            catalogue.  Measured \snrtime values are computed from detections
            of known pulsars in PALFA observations.  The increased scatter and
            bias towards higher \snr ratios of longer-period pulsars are
            consistent with reduced sensitivity to these pulses due to red
            noise (see \sect~\ref{sec:sensitivity} and
            Fig.~\ref{fig:sensitivity_curves}). Known pulsars without reported
            flux densities and uncertainties are excluded, as are pulsars that
            have reported flux densities consistent with 0~mJy. Also excluded
            from the plot are 15 known pulsars with published flux densities
            that were detected in observations pointed more than $3'$ from the
            position of the pulsar. This is because the actual beam pattern
            differs considerably from the theoretical Airy disk beam pattern
            beyond \app{$3'$}, making it difficult to reliably estimate the
            expected \snrtime.  The dashed line indicates equality of the
            expected and measured \snrtime values, and the dotted lines are at
            a factor of two above and below equality.
        \label{fig:knownpsr_snr}}
\end{figure}

By comparing expected and measured signal-to-noise ratios against
pulsar spin period we find that longer-period pulsars show an increase
scatter in \snrtime ratio as well as a bias towards larger ratios (see
Fig.~\ref{fig:knownpsr_snr}).  This is consistent with the reduced
sensitivity to long-period pulsars due to red noise we find from our
sensitivity analysis using synthetic pulsar signals (see
\sect~\ref{sec:injected}).

In addition to the 83 known pulsars with published $S_{1400}$ detected with the
PALFA \presto pipeline, there are 50 more that do not have values for $S_{1400}$
listed in the ATNF catalogue. The complete list of 128 previously discovered
pulsars blindly re-detected by the PALFA \presto pipeline is in Table
\ref{tab:redetections}.


\subsection{Known Pulsars Missed}
\label{sec:knownmissed}

In addition to the 268 detections of 128 separate known pulsars mentioned in
\sect~\ref{sec:known}, there were 7 instances in which a known pulsar was not
detected by the search pipeline, despite being detected when subsequently
folding the search data with the most recently published ephemeris. In all
cases the data were badly affected by RFI; there are strong signals within one
Fourier bin of the pulsar period. Furthermore, these are long-period pulsars,
which are more difficult to detect than expected due to red noise in the data.
It is therefore not entirely surprising that these observations did not result
in detections. A thorough analysis of the effects of RFI and red noise on the
sensitivity to long period pulsars is therefore crucial, and forms the
discussion of the following section.

\section{Assessing the Survey Sensitivity}
\label{sec:injected}
The sensitivity of pulsar observations is typically estimated using the
radiometer equation (Eq.~\ref{eq:radiometer}). In principle, the effects of DM,
period, and pulse width on sensitivity are adequately described by the
radiometer equation. The expression derived by \citet[][see their Appendix
A]{cc97}, includes a more complete description of pulse shape and the
effect of DM, which causes distortions of the pulse profile. However, neither
of these equations includes the effect of RFI. In this section, we describe a
prescription for accurately modeling the sensitivity of pulsar search
observations including the effect of RFI, as well as its dependence on period,
DM, and pulse width.

To estimate the survey sensitivity we injected synthetic pulsar signals into
actual survey data, and attempted to recover the period and DM of the input
signal using our pipeline. By using synthetic signals we can also better
determine the selection effects imposed by our pipeline.

\subsection{Constructing a Synthetic Pulsar Signal}
For this work, a simple synthetic pulsar signal was constructed for a given
combination of period, DM, phase-averaged flux density, and profile shape. 
Once the relevant parameters were chosen (see \sect~\ref{sec:injectsearch} and
Table~\ref{tab:injparams}), a two-dimensional pulse profile (intensity vs. spin
phase and observing frequency) was generated.

The pulse profile of each frequency channel was smeared by convolving with a box-car
whose phase width corresponded to the dispersion delay within the channel, as
well as scattered by convolving with a one-sided exponential function with a
characteristic phase width corresponding to the pulse broadening time scale. We
determined the scattering time scale using Eq.~\ref{eq:scattering}. Care was
taken to conserve the area under the profile during the convolutions. The
scaling factor applied to the synthetic signals was determined by
flux-calibrating the PALFA observing system (see \sect  \ref{sec:calibration}).

\subsection{Calibration}
\label{sec:calibration}
On 2013 December 21, we observed the radio galaxy 3C~138 in order
to calibrate the central beam of ALFA. Three observations using the standard
survey set-up described in \sect \ref{sec:observations} were conducted, but with
5-min integrations, and with the calibration diode being pulsed on and off at
40~Hz.  The on-source scan of 3C~138 was preceded by an off-source scan 0.5\deg
to the north of 3C~138 and followed by a similar off-source scan 0.5\deg to
the south.

The calibration observation data were converted to 4-bit samples, and the Mock
spectrometer sub-bands were combined (see \sect~\ref{sec:pipeline-preproc}).
The data were folded at the modulation frequency of the calibrator diode using
\foldpsrfits of \psrfitsutils. Next, the on-cal and off-cal levels in
    the on-source and off-source observations were used to relate the flux
    density of the calibration diode with the cataloged flux density of 3C~138
    \citep[for details, see e.g.][]{lk04}. The result is the flux density of
    the calibration diode as a function of observing frequency. In practice,
this was done using \fluxcal of
\psrchive\footnote{http://psrchive.sourceforge.net/}.

The per-channel scaling factors between flux density and the observation data
units were determined by applying the calibration solution along with the
calibration diode signal. This procedure determines the absolute level of the
injected signal corresponding to a target phase-averaged flux density, as well as the
shape of the bandpass.

\subsection{Injection Trials}
\label{sec:injectsearch}
Artificial pulsar signals were injected into the data by summing the
two-dimensional, smeared, scattered, and scaled synthetic pulse profile with
the data at regular intervals corresponding to the period of the synthetic
pulsar. The scaling was determined using the calibration procedure described in
\sect~\ref{sec:calibration}. The resulting data file, including the injected
signal, was written out with 32-bit floating-point samples in \sigproc
``filterbank'' format\footnote{http://sigproc.sourceforge.net/} to avoid having
to quantize the weak synthetic pulsar signal. 

\begin{figure}[t]
        \includegraphics[width=\columnwidth]{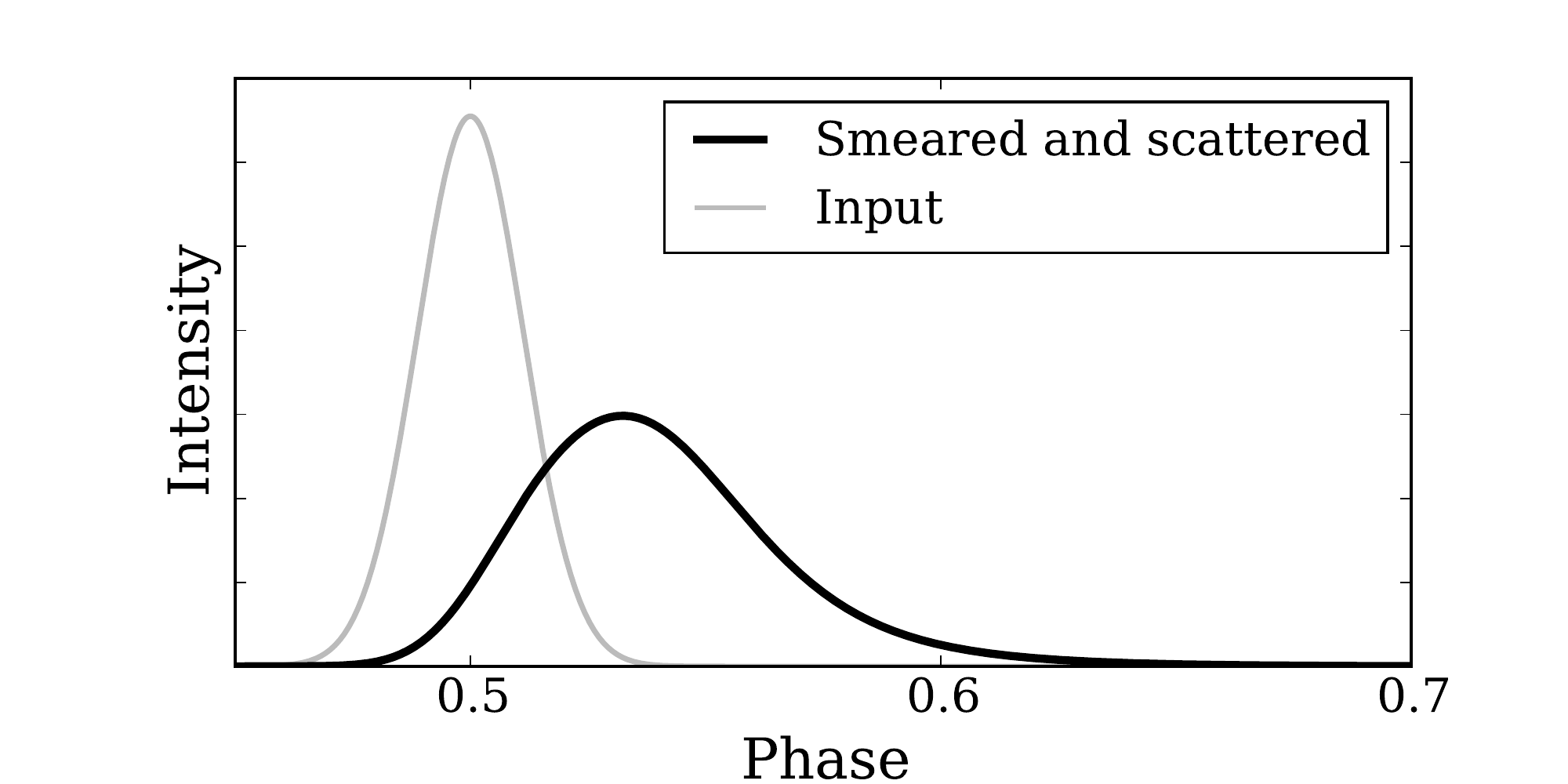}
        \caption{The profile of a synthetic $P=5$\,ms pulsar consisting of a
            single von~Mises component with FWHM=2.6\% (gray), and the same profile broadened
            according to DM=250\,\dmunit. The broadening is caused by dispersive
            smearing within each channel and scattering according to
            Eq.~\ref{eq:scattering}.  Note that the plot is zoomed into the
            region: $0.45 < \phi < 0.7$.
        \label{fig:inject_example}}
\end{figure}

Many synthetic signals with a broad range of parameters were required to build
a comprehensive picture of the survey sensitivity (see
Table~\ref{tab:injparams}).  In total, 17 periods were selected between
0.77\,ms and 11\,s along with six DMs ranging from 10 to 600\,\dmunit. In all
cases, the profile of the synthetic signal was chosen to have a single centered
von~Mises component with a FWHM selected from 5 possible values between
$\sim$1.5\,\% and $\sim$24\,\% of the period. The example profile in
Figure~\ref{fig:inject_example} shows the case where FWHM=2.6\,\%.  The
synthetic signals were injected into 12 different observations to determine the
survey sensitivity in a variety of RFI conditions.
All 12 observations used in this analysis are from late 2013 and from the
central beam of ALFA.  Although the gains of the outer beams are lower than
that of the central beam, the response of the observing system and pulsar
search pipeline to RFI and red noise derived for the central beam should also
apply to the outer beams.

The total number of combinations of synthetic signals and observations is
$>$\,6000. Multiple trials, each with a different amplitude, were 
constructed, injected, and searched to determine the sensitivity limit at each
point in (period, DM, pulse FWHM) phase-space. To reduce the computational burden,
not all possible combinations of parameters were used. In particular, only the
profile with FWHM~$\sim$3\,\% was injected into all 12 observations. The
remaining four profiles shapes were only injected into a single observation. This
still permits the determination of the dependence of $S_\mathrm{min}$ on pulse width.

\subsection{Realistic Survey Sensitivity}
\label{sec:sensitivity}
\begin{figure*}[t]
        \includegraphics[width=\textwidth]{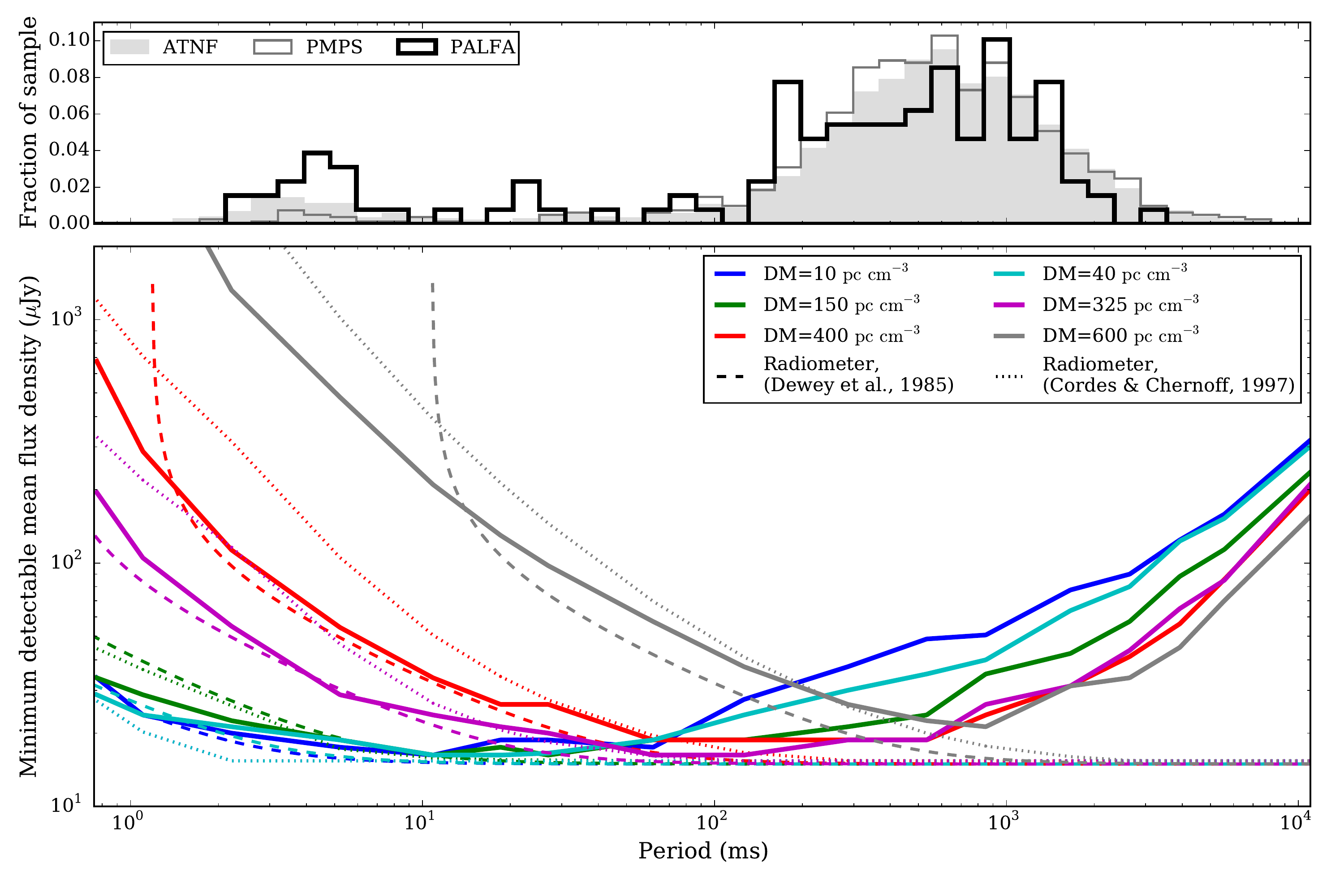}
        \caption{
            \textit{Top} -- Period distribution of all Galactic radio pulsars,
            excluding RRATs, listed in the ATNF catalogue, as well as those
            found in the PALFA and Parkes Multibeam Pulsar Survey (PMPS). \\
            \textit{Bottom} -- Minimum detectable phase-averaged flux density curves
            for the PALFA survey as determined using synthetic pulsar signals
            with FWHM=2.6\,\%. The reduction in sensitivity at long periods is
            due to red noise in the data. We see clear discrepancies
            when comparing the measured curves with the analogous sensitivity
            limits derived with the commonly used radiometer equation
            \citep{dtws85}.  Sensitivity to long-period pulsars is
            overestimated, and sensitivity to MSPs is underestimated. However,
            the formulation of the radiometer equation by \citet{cc97} is more
            complete -- albeit less frequently used -- and better
            models the sensitivity in the short-period regime. See
            \sect~\ref{sec:sensitivity}.
        \label{fig:sensitivity_curves}}
\end{figure*}

It is well known \citep{dtws85} that the minimum detectable flux density of a
pulsar depends on the intrinsic width of its profile, as well as the DM,
because dispersive smearing and scattering broaden the profile. It is also
reasonable to expect a reduction of sensitivity due to RFI and red noise, even
with the red noise suppression algorithms employed (see
\sect~\ref{sec:deredden}).
By recovering injected signals using the pipeline described in
\sect~\ref{sec:analysis}, we have determined the true sensitivity of the PALFA
survey, and its dependence on spin period and DM (see
Fig.~\ref{fig:sensitivity_curves}).   
We found the commonly used version of the radiometer equation
\citep[Eq.~\ref{eq:radiometer};][]{dtws85} overestimates the survey sensitivity
to long-period pulsars. For example, for $P = 0.1$--2.0\,s pulsars with DM $>
150$\,\dmunit (the majority of the pulsars we expect to find with PALFA), the
degradation in sensitivity compared with the ideal case is a factor of
$\sim$1.1--2.

We have also confirmed the claim by \citet{cc97} that the \citet{dtws85}
radiometer equation underestimates the sensitivity to high-DM MSPs, by not
correctly modeling the distortion of the profile due to smearing and
scattering. The more accurate variant of the radiometer equation from
\citet{cc97} better matches our measured sensitivity curves in the MSP regime,
thanks to its inclusion of the profile shape and distortions. However, the
degraded sensitivity we find at long periods is still not properly modeled with
these adjustments.

Red noise present in pulsar search data due to RFI, receiver gain fluctuations,
and opacity variations of the atmosphere makes it difficult to detect
long-period radio pulsars. Our analysis has shown that for the PALFA survey, at
low DMs, the reduction in sensitivity already affects pulsars with periods of
$\sim\!100$\,ms. Fortunately, the effect is slightly less significant for
pulsars with higher DMs. This is evident in
Figure~\ref{fig:sensitivity_curves}.

\begin{figure*}[t]
        \includegraphics[width=\textwidth]{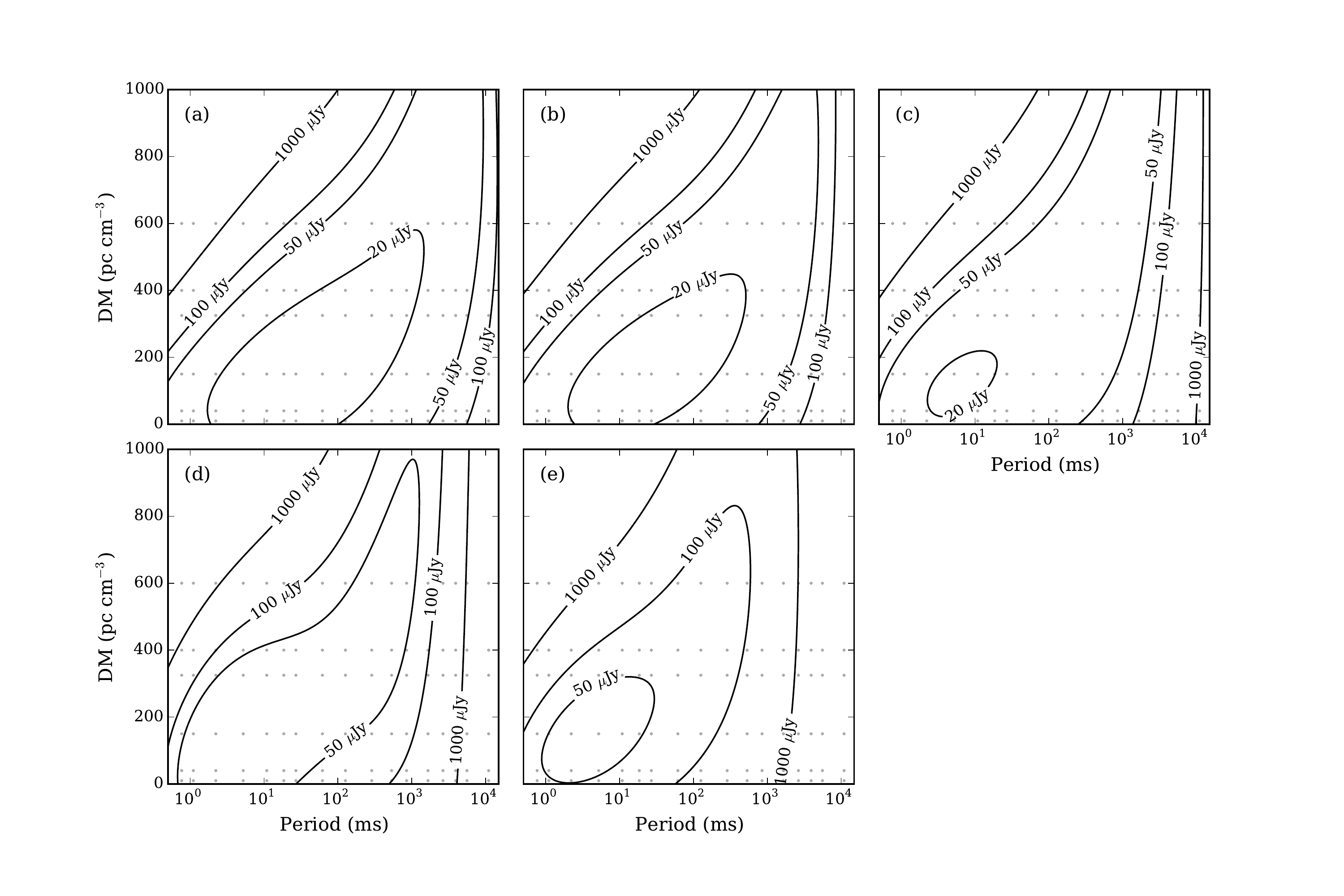}
        \caption{PALFA survey sensitivity as a function of DM and spin period.
                The maps are determined using synthetic pulsar signals injected
                into observations and recovered using the pipeline. Contours
                correspond to minimum detectable phase-averaged flux densities
                of 20, 50, 100, 1000\,$\mu$Jy. The five panels (a)--(e)
                correspond to profile FWHMs of 1.5, 2.6, 5.9, 11.9, 24.3\,\%,
                respectively. In all cases, the profile consists of a single
                centered von Mises component (see Fig.~\ref{fig:inject_example}
                for an example). The period, DM combinations used in the
                sensitivity analysis are shown with the small dots.
        \label{fig:sensitivity_maps}}
\end{figure*}

We have parameterized the sensitivity curves by fitting $\log S_\mathrm{min}$
vs. DM with a quadratic function and modeling how these curves depend on
period. To estimate $S_\mathrm{min}$ at an arbitrary profile width, we first
estimate $S_\mathrm{min}$ at each of the five trial widths, then fit a
quadratic function in $\log S_\mathrm{min}$ vs.  width, and use the parameters
of the fit to calculate $S_\mathrm{min}$ at the desired width.  This ad-hoc
scheme provides reliable estimates of $S_\mathrm{min}$ within the intervals
used for trial values of period, DM, and width, as well as for modest
extrapolation.  
Sensitivity maps for each of the five profile widths used are shown in
Figure~\ref{fig:sensitivity_maps}. 

\section{Population Synthesis Analysis}
\label{sec:popsynth}
We have used the sensitivity curves determined above (see
\sect~\ref{sec:sensitivity}) to re-evaluate the expected yield of the PALFA
survey by performing a population synthesis analysis with
\psrpoppy\footnote{https://github.com/samb8s/PsrPopPy} \citep{blrs14}. 

Galactic populations of non-recycled pulsars were simulated using the radial
distribution from \citet{lfl+06} and a Gaussian distribution of heights
above/below the plane with a scale height of 330\,pc. The pulsar periods were
described by a log-normal distribution with $\left < \log P \right > = 2.7$ and
$\sigma_{\log P} = -0.34$ \citep{lfl+06}. The pulse-width-to-period
relationship was also taken from \citet{lfl+06}. We used a log-normal
luminosity distribution described by the best-fit parameters found by
\citet{fk06}, $\left < \log L \right > = -1.1$ and $\sigma_{\log L} = 0.9$.

We created 5000 simulated pulsar populations, each containing enough pulsars
such that a simulated version of the Parkes multi-beam surveys detected 1038 pulsars, 
the number of non-recycled pulsars detected by the actual surveys. We then
compared the pulsars in each of these populations against a list of PALFA
observations\footnote{For each observation we used the sky position,
    integration time, zenith angle and beam number. We used the model of gain
and system temperature dependence on zenith angle provided by the observatory.
We assumed the six outer beams have a gain of $\sim$80\,\% of the central beam,
consistent with the gains reported by \citet{cfl+06}.}, and estimated their
significance using
the radiometer
equation. Pulsars with $(S/N)_\mathrm{expect} > 11.3$ were considered
detected\footnote{The value of $(S/N)_\mathrm{expect}$ was chosen such that the
minimum detectable flux density coincided with the measured sensitivity
curves for a duty cycle of 2.6\%.}. Next, we compared the flux-density for each ``detected''
    pulsar against the parameterized PALFA sensitivity curves to determine if
    the pulsar also has a sufficiently large flux density to lie above the
    measured sensitivity curves. For each pulsar, the measured sensitivity
curves are shifted according to the zenith angle of the observation, the gain
of the beam used, the sky temperature and the angular offset between the pulsar
position and the beam center.

\begin{figure}[t]
        \includegraphics[width=\columnwidth]{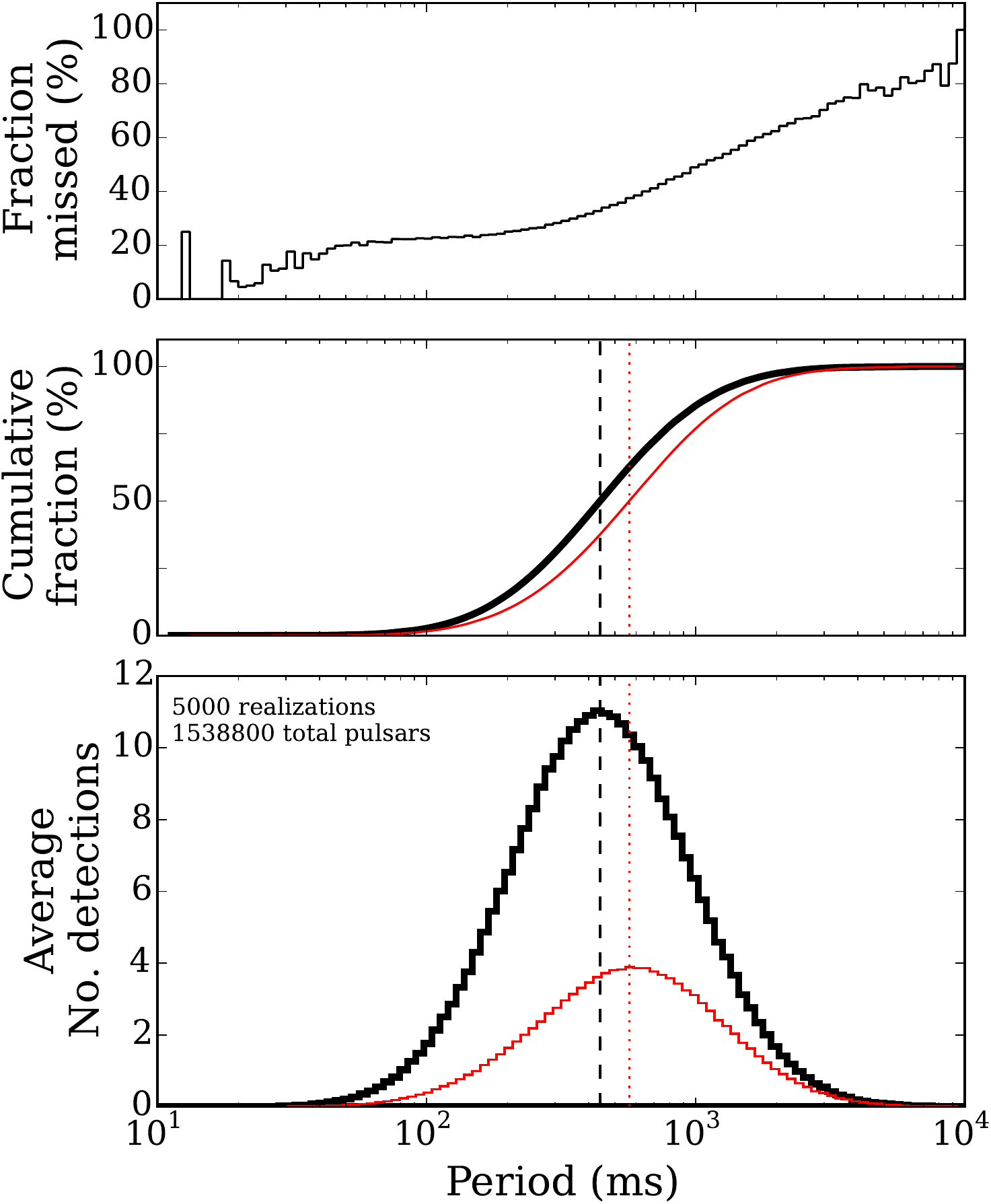}
        \caption{
        \textit{Top} -- Fraction of potentially detectable pulsars missed by
        PALFA due to red noise as a function of spin period, assuming
        the underlying pulsar population is accurately modeled by our input
        distributions (i.e. the distributions in \citet{lfl+06}, see
        \sect~\ref{sec:popsynth}). \\
        \textit{Middle} -- Cumulative fraction of simulated pulsars (thick
        black line), and pulsars missed (thin red line) as a function of pulse
        period. \\
        \textit{Bottom} -- Period distribution of potentially detectable
        simulated population of un-recycled pulsars averaged over 5000
        realizations (thick black line) compared with the period distribution
        of pulsars expected to be missed due to red noise (thin red line). The
    median spin period of the potential detectable pulsars \nts{($P\simeq440$\,ms)} is
        shown by the dashed black line, and the median spin period \nts{($P\simeq566$\,ms)}
        of the missed pulsars is shown by the dotted red line.
        \label{fig:popsynth-pdist}
        }
\end{figure}

We found \nts{$35 \pm 3$\,\%} of the simulated pulsars having fluxes
above the theoretical sensitivity threshold derived from the radiometer equation
(Eq.~\ref{eq:radiometer}) are not sufficiently bright to be ``detected'' by our
measured sensitivity limits for the PALFA survey (e.g.
Fig.~\ref{fig:sensitivity_maps}) due to the residual effect of red noise and
RFI following the extensive mitigation procedures described in
\sect~\ref{sec:pipeline-rfi}. The median period of the pulsars missed is
\nts{$P_\mathrm{miss} \simeq 566$\,ms}, which is considerably longer than the
median period of the potentially detectable pulsars brighter than the
radiometer-equation-based threshold, \nts{$P_\mathrm{det.} \simeq 440$\,ms}
(see Fig.~\ref{fig:popsynth-pdist}).  

Our 5000 realizations of simulated Galactic pulsar populations, adjusted for
the reduced sensitivity to long-period pulsars, suggest \nts{$217 \pm 15$}
un-recycled pulsars should be detected in PALFA Mock spectrometer observations,
given the current processed pointing list.  As of 2015 January, 
\nts{241} un-recycled pulsars have been discovered/detected in
PALFA observations with the Mock spectrometers. 

The number of un-recycled pulsar detections predicted for the PALFA survey by
\citet{slm+14} is an overestimate for two reasons. First, their analysis used a
threshold \snr$=9$. Given the observing parameters assumed, a more appropriate
threshold of \snr$=11.3$ should have been used to correspond to the minimum
detectable flux density we find ($S_\mathrm{min}=0.015$\,mJy). Second, the
analysis by \citet{slm+14} did not include the effect of red noise, which we
have shown reduces the number of pulsars expected to be found in the PALFA
survey by 35\%.

\section{Discussion}
\label{sec:discussion}

The detailed sensitivity analysis of \sect~\ref{sec:sensitivity} confirms that, on 
average, the PALFA survey is as sensitive to MSPs and mildly recycled pulsars
as expected from the radiometer equation. However, the survey is less sensitive to
long-period pulsars than predicted. The degradation in sensitivity is
    between 10\% and a factor of 2 for the majority of pulsars we expect to
    find in the PALFA survey (spin periods between 0.1\,s and 2\,s and DM $>
    150$\,\dmunit), and up to a factor of $\sim10$ in the worst case (DM $<
    100$\,\dmunit and P $> 2$\,s; this fortunately corresponds to a parameter
    space that contains far fewer expected pulsars). The reduction of
    sensitivity is likely caused by red noise present in the observations.

The empirical sensitivity curves we determined apply
specifically to the PALFA survey, its observing set-up, and the search
algorithms used. Because the effects of red noise on radio pulsar survey
sensitivity have the potential to be significant, as in the case of PALFA, we
strongly suggest measuring the impact of red noise on other surveys by
performing similar analyses to what we described in \sect~\ref{sec:injected}.
Also, future population analyses should include these measured effects of red
noise rather than assuming the theoretical radiometer equation
\citep[e.g.][]{lfl+06,fk06} when deriving spatial, spin, and luminosity
distributions for the underlying Galactic population of pulsars.

What are the potential ramifications of reduced sensitivity to long-period
pulsars being unaccounted for in population synthesis analyses? First, the
existence of radio-loud pulsars beyond the ``death line'' is important to our
understanding of the radio emission mechanism in pulsars. For example, the
existence of the 8.5-s PSR~J2144$-$3933 contradicted several existing emission
theories \citep[][]{ymj99,zhm00}. 
The existence of a larger population of slowly rotating pulsars, particularly
the discovery of pulsars so slow that existing theories cannot explain their
radio emission, would further constrain models.

It is also possible there is a larger population of highly magnetized
rotation-powered pulsars and quiescent radio-loud magnetars that have been
missed by the lower than predicted sensitivity of pulsar surveys.  Radio
emission from three of the four known radio-loud magnetars was detected
following high-energy radiative events \citep{crh+06,crhr07,sj13,efk+13}.
However, the other radio-loud magnetar PSR~J1622$-$4950 was discovered from its
radio emission \citep{lbb+10,ok14}.  There is no evidence that the turn-on of
PSR~J1622$-$4950 at radio wavelengths was preceded by a high-energy event.  The
possibility that radio emission from magnetars is not always accompanied by
X-ray or $\gamma$-ray emission means it is crucial to understand the biases
against finding such long-period pulsars.  Characterizing, and hopefully
uncovering a hidden population of radio-loud magnetars, as well as highly
magnetized-rotation powered pulsars, will help clarify the relationship between
these two classes of pulsars, as well as the influence of strong magnetic
fields on emission properties (e.g flux and spectral index variability).

It may be possible to address the reduced sensitivity to
long-period pulsars by utilizing algorithms that perform better in the presence
of red noise, as well as algorithms that remove red noise without suppressing
the pulsar signal. 

Long-period pulsars may be found via their harmonics even if red noise obscures the
signal in the Fourier domain at the fundamental frequency of the pulsar. The
detection of the pulsar signal will be reduced in two ways. First, the harmonic
summing algorithm will exclude the power contained at the fundamental and low
harmonic frequencies, which can contain large amounts of power, especially in
the case of pulsars with wide profiles. Second, by not being based at the
fundamental frequency of the pulsar, the harmonic summing algorithm will skip
slower, more significant harmonics in favor of weaker harmonics at higher
frequencies. Despite the reduction in sensitivity several pulsars have
been found in the PALFA survey thanks to their higher harmonic content.

One suggested method of improving sensitivity to long-period pulsars is by
using the Fast-folding algorithm \citep[FFA; see e.g.][and references
therein]{lk04,kml+09}. The periodograms produced by the FFA, a
time-domain algorithm, are generated from computing a significance metric
from pulse profiles. Thus, the broad profile features caused by red noise
pose a problem for FFA-based searches. In short, the FFA is not immune to the
degradation of sensitivity to long-period pulsars described above. However it
does have the advantage of coherently summing \textit{all} harmonics of a given
period and greater period resolution than the DFT. These two factors should
make the FFA slightly more sensitive to long-period pulsars, especially those
with narrow profiles, than the Fourier Transform techniques described in
\sect~\ref{sec:periodicity}, which is limited in the number of harmonics that
can be summed \citep[typically incoherently;][]{kml+09}. The FFA has only been
used sparingly in large-scale pulsar searches \citep[e.g.][]{kml+09}. A more
systematic investigation and application of the FFA is warranted.

Another algorithm that might have better performance in the presence of red
noise is the single-pulse search technique described in
\sect~\ref{sec:singlepulse}. Single-pulse search algorithms are known to be
more sensitive than standard FFT techniques to long-period pulsars in short
observations \citep{dcm+09,kkl+15}. This is because of the natural variability
of pulsar pulses and small number of pulses. Pulse-to-pulse variability was not
included in the synthetic pulsar signals used in our sensitivity analysis and
no single pulse searching was performed. It is likely that the
sensitivity curves determined in this work are partially compensated by the
single-pulse search techniques already in place, especially considering the
recent suggestion that pulsars with $P > 200$\,ms have a greater likelihood of
being detected in single-pulse searches than faster pulsars \citep{kkl+15}.
However, the extent of this compensation depends on the pulse-energy
distributions of pulsars and the relative significances of their detections
in periodicity and single-pulse searches.

\section{Conclusions}
\label{sec:conclusion}

We described the \presto-based PALFA pipeline, the primary data analysis
pipeline used to search PALFA observations made with the Mock spectrometers.
This pipeline has led to the discovery of 41 periodicity pulsars and 5 RRATs,
the re-detection of \nts{60} pulsars previously discovered in the survey (using
other pipelines), and the detection of \nts{128} previously known pulsars. The
\presto-based pipeline described here consists of several complementary search
algorithms and RFI-mitigation strategies. The performance of the pipeline was
determined by injecting synthetic pulses into actual survey observations and
recovering the signals.

We have found that the PALFA survey is as sensitive to fast-spinning pulsars as
expected by the theoretical radiometer equation. However, in the case of
long-period pulsars, we have found that there is a reduction in the sensitivity
due to RFI and red noise in the observations. The actual detection threshold
for pulsars with $P>4$\,s at $\mathrm{DM} < 150$\,\dmunit is up to \app{10}
times higher than predicted by the theoretical radiometer equation.  We have
performed a population synthesis analysis using this empirical model of the
survey sensitivity. Our analysis indicates that \nts{$35 \pm 3$\,\%} of
pulsars, with predominantly long periods, are missed by PALFA, compared to
expectations based on theoretical sensitivity curves derived using the
radiometer equation.

The magnitude of the effect of red noise on the PALFA survey's sensitivity to
long-period pulsars is surprising and should be taken into account in future
population synthesis analyses. Furthermore, the effect of red noise on other
radio pulsar surveys should be quantified in a similar manner and be included
in population synthesis analyses to ensure the distributions determined for the
underlying pulsar population are robust. The presence of more long-period
pulsars could have implications on the location of the pulsar death line, the
structure of pulsar magnetospheres and radio emission mechanism, as well as the
relationship between canonical pulsars, highly magnetized rotation-powered
pulsars, radio-loud magnetars, and RRATs.

\acknowledgements
The Arecibo Observatory is operated by SRI International under a cooperative
agreement with the National Science Foundation (AST-1100968), and in alliance
with Ana G. M\'endez-Universidad Metropolitana, and the Universities Space
Research Association. The CyberSKA project was funded by a CANARIE NEP-2 grant.

Computations were made on the supercomputer Guillimin at McGill University,
managed by Calcul Qu\'{e}bec and Compute Canada. The operation of this
supercomputer is funded by the Canada Foundation for Innovation (CFI),
NanoQu\'{e}bec, RMGA and the Fonds de recherche du Qu\'{e}bec - Nature et technologies
(FRQ-NT).

We would like to recognize the help of Bryan Fong in developing the decision
tree AI, Mark Tan for his contributions to the decision tree AI and Survey
Diagnostics CyberSKA application, and the Sequence Factory for developing of
the CyberSKA-integrated PALFA applications. PL would like to thank David
Champion for helpful discussions.

PL acknowledges the support of IMPRS Bonn/Cologne and FQRNT~B2.
PALFA work at Cornell University is supported by NSF grant PHY~1104617. 
VMK receives support from an NSERC Discovery Grant and Accelerator
Supplement, Centre de Recherche en Astrophysique du Qu\'{e}bec, an R.~Howard
Webster Foundation Fellowship from the Canadian Institute for Advanced Study,
the Canada Research Chairs Program and the Lorne Trottier Chair in Astrophysics
and Cosmology.
JWTH acknowledges funding from an NWO Vidi fellowship and ERC Starting Grant
``DRAGNET'' (337062).
PCCF and LGS gratefully acknowledge financial support
by the European Research Council for the ERC Starting Grant
BEACON under contract no. 279702.
Pulsar research at UBC is supported by an NSERC Discovery Grant and Discovery
Accelerator Supplement and by the Canadian Institute for Advanced Research.

\clearpage

\begin{table}
\centering
\caption{PALFA Mock Spectrometer Observing Set-up Parameters \label{tab:obs_setup}}
\begin{tabular}{lc}
    \hline\hline\\[-2mm]
    Parameter & Value \\[1mm]
    \hline\\[-2mm]
    \multicolumn{2}{c}{General} \\
    \hline \\[-2mm]
    Sample Time, $\tau_\mathrm{samp}$ ($\mu$s) & 65.476 \\ 
    Integration Time$^a$, $t_\mathrm{obs}$ (s) & 268 (Inner Galaxy, 32\deg \lapp~$l$ \lapp~77\deg) \\
                        & 180 (Anti-Center, 168\deg \lapp~$l$ \lapp~214\deg) \\
    \hline \\[-2mm]
    \multicolumn{2}{c}{High Sub-Band} \\
    \hline \\[-2mm]
    Number of Channels & 512 \\
    Low Frequency (MHz) & 1364.290 \\ 
   High Frequency (MHz) & 1536.016 \\ 
    \hline \\[-2mm]
    \multicolumn{2}{c}{Low Sub-Band} \\
    \hline \\[-2mm]
    Number of Channels & 512 \\
    Low Frequency (MHz) & 1214.290 \\ 
   High Frequency (MHz) & 1386.016 \\ 
    \hline \\[-2mm]
    \multicolumn{2}{c}{Merged Band} \\
    \hline \\[-2mm]
    Number of Channels & 960 \\
         Low Frequency (MHz) & 1214.290 \\ 
         Center Frequency (MHz) & 1375.489 \\
        High Frequency (MHz) & 1536.688 \\ 
 Bandwidth, $\Delta f$ (MHz) & \phantom{1}322.398 \\
        Channel Bandwidth, $\Delta f_\mathrm{chan}$ (kHz) & \phantom{1}335.831 \\
    \hline \\[-2mm]
\end{tabular}
\\ \footnotesize \raggedright
$^a$This includes the $\sim5-10$\,s when the calibration diode is
    turned on, which is not usable for searching for pulsars. The interval of
    the observation containing this calibration signal is removed prior to our
    analysis (see \sect~\ref{sec:pipeline-preproc}).
\end{table}

\begin{table}
\centering
\caption{Breakdown of PALFA Mock Spectrometer Data \label{tab:data}}
\begin{tabular}{cccccc}
    \hline\hline\\[-2mm]
    & No. Beams$^a$ & No. Unique & Sky Coverage & 
                Completeness$^b$, $|b| < $2\deg & Completeness$^b$, $|b| < $5\deg \\
                & & Sky Positions & (\sqdeg) & (\%) & (\%) \\
    \hline\\[-2mm]
    \multicolumn{6}{c}{Inner Galaxy (32\deg \lapp~$l$ \lapp~77\deg)} \\
    \hline \\[-2mm]
    Observed & 40705 & 38479 & 94 & 69 & 32 \\ 
    Archived & 35030 & 33243 & 81 & 60 & 27 \\ 
    Analyzed & 33888 & 32499 & 80 & 58 & 27 \\ 
    \hline \\[-2mm]
    \multicolumn{6}{c}{Anti-Center (168\deg \lapp~$l$ \lapp~214\deg)} \\
    \hline \\[-2mm]
    Observed & 60305 & 26194 & 64 & 30 & 18 \\ 
    Archived & 52659 & 21990 & 54 & 23 & 15 \\ 
    Analyzed & 51445 & 21899 & 54 & 23 & 15 \\ 
    \hline\\[-2mm]
\end{tabular}
\\ \footnotesize \raggedright
%
%
$^a$There are 7 beams per pointing.\\
$^b$The completeness percentages are relative to the number of pointings we
will eventually cover with the Mock spectrometers.\\
\end{table}

\begin{table}
\centering
\caption{Dedispersion Plan for Mock Spectrometer Data \label{tab:ddplan}}
\begin{tabular}{X{-1} cccccc}
    \hline\hline\\[-2mm]
    \hdr{DM range} & DM step size & No. DMs & 
               No. sub-bands & Sub-band DM spacing &
               Down-sample factor & Approx. Computing\\
                  \hdr{(\dmunit)} & (\dmunit) & & & (\dmunit) & & (\%)\\
    \hline\\[-2mm]
           0   -  212.8 & 0.1 & 2128 & 96 &   7.6 &  1 & 73.19 \\
         212.8 -  443.2 & 0.3 &  768 & 96 &  19.2 &  2 & 12.20 \\
         443.2 -  534.4 & 0.3 &  304 & 96 &  22.8 &  3 & 8.13 \\
         534.4 -  876.4 & 0.5 &  684 & 96 &  38.0 &  5 & 2.93 \\
         876.4 -  990.4 & 0.5 &  228 & 96 &  38.0 &  6 & 2.44 \\
         990.4 - 1826.4 & 1.0 &  836 & 96 &  76.0 & 10 & 0.73 \\
        1826.4 - 3266.4 & 2.0 &  720 & 96 & 144.0 & 15 & 0.24 \\
        3266.4 - 5546.4 & 3.0 &  760 & 96 & 228.0 & 30 & 0.08 \\
        5546.4 - 9866.4 & 5.0 &  864 & 96 & 360.0 & 30 & 0.05 \\ 
    \hline\\[-2mm]
\end{tabular}
\\ \footnotesize \raggedright
\textsc{Note.} --- See also Fig.~\ref{fig:ddplan} for the pulse
    broadening as a function of DM due to dispersive
    smearing and this dedispersion plan.\\ 
\end{table}

\begin{table}
\centering
    \caption{Heuristic Candidate Ratings \label{tab:ratings}}
\begin{tabular}{ll}
    \hline\hline\\[-2mm]
    Rating & Description\\
    \hline\\[-2mm]
    \multicolumn{2}{c}{Profile Ratings$^a$} \\
    \hline \\[-2mm]
    Duty Cycle & Fraction of profile bins larger than half the maximum value of the profile \\
 Peak over RMS & Maximum value of the profile divided by the RMS \\
    \hline \\[-2mm]
    \multicolumn{2}{c}{Profile Ratings (Gaussian Fitting)$^a$} \\
    \hline \\[-2mm]
           Amplitude & Amplitude of a single Gaussian component fit to the profile \\
Single Component GoF & Goodness of Fit of a single Gaussian component fit to 
                    the profile \\
                FWHM & Full-width at half-maximum of a single Gaussian component 
                    fit to the profile \\
      No. Components & Number of Gaussian components required to acceptably fit 
                    the profile \\
                    & (up to 5 components) \\
 Multi-Component GoF & Goodness of fit of the multiple Gaussian component fit 
                    (up to 5 components) \\
         Pulse Width & Ratio of narrowest component of the multiple Gaussian fit 
                    compared to the \\
                    & pulse broadening (excluding scattering) \\
    \hline \\[-2mm]
    \multicolumn{2}{c}{Time vs. Phase Ratings} \\
    \hline \\[-2mm]
    Period Stability & Fraction of good time intervals that deviate in phase by $\leq0.02$ \\
    Frac. of Good Sub-ints & Fraction of time intervals that contain the pulsar signal \\
  Sub-int. SNR Variability & The standard deviation of sub-integration \snrs\\
    \hline \\[-2mm]
    \multicolumn{2}{c}{Frequency vs. Phase Ratings} \\
    \hline \\[-2mm]
    Frac. of Good Sub-bands & Fraction of sub-bands that contain the pulsar signal \\
   Sub-band SNR Variability & The standard deviation of sub-band \snrs \\
    \hline \\[-2mm]
    \multicolumn{2}{c}{DM Ratings} \\
    \hline \\[-2mm]
    DM Comparison & Ratio of the standard deviation of the profile at DM=0\,\dmunit \\
    (standard deviation) & and at the optimal DM \\
            DM Comparison ($\chi^2$) & Ratio of the $\chi^2$ of the profile at DM=0\,\dmunit and at the optimal DM \\
                  DM Comparison (peak) & Ratio of the peak value of the profile at DM=0\,\dmunit and at the optimal DM \\
    \hline \\[-2mm]
    \multicolumn{2}{c}{Miscellaneous Ratings} \\
    \hline \\[-2mm]
    Known Pulsar & A measure of how similar the candidate period and DM are to a nearby pulsar \\
                 & (also checks harmonic relationships) \\
       Mains RFI & A measure of how close the topocentric frequency is to 60\,Hz, or a harmonic \\
      Beam Count & The number of beams from the same pointing containing another candidate \\
                 & with the same period \\
    \hline\\[-2mm]
\end{tabular}
\\ \footnotesize \raggedright
\textsc{Note.} --- See \sect~\ref{sec:ratings} for more
details on how ratings are used to select candidates.\\
    $^a$Prior to computing ratings, the profile is normalized such that median
    level is 0 and the standard deviation is 1.\\
\end{table}

\begin{table}
\centering
    \caption{Pulsars Discovered in Mock Spectrometer Data with the \presto 
    Pipeline \label{tab:discoveries}}
\begin{tabular}{l d{4.2} d{4.1} d{2.2} d{1.6}}
    \hline\hline\\[-2mm]
    Name & \hdr{Disc. Period} & \hdr{Disc. DM} &
               \hdr{Disc. Significance} & \hdr{Flux Density$^a$} \\
                 & \hdr{(ms)} & \hdr{(\dmunit)} & \hdr{(\sigfourier)} & \hdr{(mJy)}\\
    \hline\\[-2mm]
    J0557+1550$^b$ &    2.55 &   102.7 &   8.34 & 0.050(6)^c                   \\
    J1850+0242$^b$ &    4.48 &   540.5 &  13.08 & 0.33                         \\
    J1851+0232     &  344.02 &   605.4 &  10.82 & 0.09                         \\
      J1853+03     &  585.53 &   290.2 &  14.28 & \hdr{--$^d$\phantom{11111}}  \\
      J1854+00$^e$ &  767.33 &   532.9 &  10.44 & \hdr{--$^d$\phantom{11111}}  \\
      J1858+02     &  197.65 &   492.1 &  14.91 & \hdr{--$^d$\phantom{11111}}  \\
    J1901+0235$^e$ &  885.24 &   403.0 &  26.7  & \hdr{--$^d$\phantom{11111}}  \\
    J1901+0300$^b$ &    7.79 &   253.7 &  11.8  & 0.113(4)^c                   \\
    J1901+0459     &  877.06 &  1103.6 &  10.93 & 0.10                         \\
      J1902+02$^e$ &  415.32 &   281.2 &   7.58 & \hdr{--$^d$\phantom{11111}}  \\
    J1903+0415$^e$ & 1151.39 &   473.5 &  12.48 & \hdr{--$^d$\phantom{11111}}  \\
    J1904+0451$^b$ &    6.09 &   183.1 &   8.78 & 0.117(9)^c                   \\
    J1906+0055     &    2.79 &   126.9 &  16.47 & 0.12                         \\
    J1906+0725     & 1536.51 &   480.4 &   7.13 & 0.05                         \\
    J1907+0256     &  618.77 &   250.4 &  12.07 & 0.19                         \\
      J1907+05     &  168.68 &   456.7 &  10.0  & \hdr{--$^d$\phantom{11111}}  \\
    J1909+1148     &  448.95 &   201.9 &  15.93 & 0.06                         \\
    J1910+1027     &  531.47 &   705.7 &   9.29 & 0.06                         \\
      J1911+09     &  273.71 &   334.7 &   7.13 & \hdr{--$^d$\phantom{11111}}  \\
      J1911+10     &  190.89 &   446.2 &   7.48 & \hdr{--$^d$\phantom{11111}}  \\
    J1913+0617     &    5.03 &   155.8 &   9.81 & \hdr{--$^d$\phantom{11111}}  \\
    J1913+1103     &  923.91 &   628.9 &   9.86 & 0.09                         \\
    J1914+0659     &   18.51 &   224.7 &  12.66 & 0.33                         \\
    J1915+1144     &  173.65 &   338.3 &  23.59 & 0.08                         \\
    J1915+1149     &  100.04 &   702.1 &   7.58 & \hdr{--$^d$\phantom{11111}}  \\
    J1918+1310     &  856.74 &   247.4 &   6.56 & \hdr{--$^d$\phantom{11111}}  \\
      J1921+16     &  936.43 &   204.7 &   8.13 & \hdr{--$^d$\phantom{11111}}  \\
    J1924+1628$^e$ &  375.09 &   542.9 &  21.12 & 0.09                         \\
      J1924+17     &  758.43 &   527.4 &  10.66 & \hdr{--$^d$\phantom{11111}}  \\ 
    J1925+1721     &   75.66 &   223.7 &  16.06 & 0.09                         \\
    J1926+1613$^e$ &  308.30 &    32.9 &  14.9  & \hdr{--$^d$\phantom{11111}}  \\
      J1930+14$^e$ &  425.71 &   209.2 &  12.15 & 0.04                         \\ 
    J1930+1723$^e$ & 1609.72 &   231.7 &   9.68 & 0.12                         \\ 
    J1931+1440     & 1779.23 &   239.3 &  23.63 & 0.12                         \\ 
      J1932+17$^e$ &   41.82 &    53.2 &  12.89 & \hdr{--$^d$\phantom{11111}}  \\
    J1933+1726     &   21.51 &   156.6 &   7.28 & 0.04                         \\ 
      J1934+19     &  230.99 &    97.6 &  18.67 & 0.10                         \\ 
      J1936+20     & 1390.88 &   205.1 &   6.6  & \hdr{--$^d$\phantom{11111}}  \\
    J1938+2012$^e$ &    2.63 &   237.1 &   8.55 & 0.02                         \\ 
    J1940+2246     &  258.89 &   218.1 &  14.47 & 0.09                         \\ 
    J1957+2516     &    3.96 &    44.0 &   6.61 & 0.04                         \\ 
    \hline\\[-2mm]
\end{tabular}
\\ \footnotesize \raggedright
    $^a$Phase-averaged flux density. Determined using the
    radiometer equation (see \sect~\ref{sec:faint}) unless otherwise noted.\\
    $^b$Pulsar was previously published by \citet{skl+15}.\\
    $^c$Flux calibrated using noise diode. Value from \citet{skl+15}.\\
    $^d$Refined position not available. Flux density could not be estimated.\\
    $^e$Pulsar was first identified using the PICS machine
                      learning candidate selection system described
                      in \sect~\ref{sec:ai}.\\
\end{table}

\clearpage
\begin{center}
    \begin{longtable}{l d{4.2} d{4.2} d{2.4} d{4.1} d{2.3}}
    \caption{\\Known Pulsars Re-detected in Mock Spectrometer Data with the
    \presto Pipeline \label{tab:redetections}} \\
    \hline\hline\\[-2 mm]
    Name & \hdr{Period} & \hdr{DM} & \hdr{ATNF $S_{1400}$} & \hdr{Measured \snr} & \hdr{Measured $S_{1400}$} \\ 
         & \hdr{(ms)} & \hdr{(\dmunit)} & \hdr{(mJy)} & & \hdr{(mJy)} \\[1mm]
    \hline\\[-2mm]
    \endhead
    \hline \\[-2mm]Continued...
    \endfoot
    \hline
    \endlastfoot
    B1848+04     &       284.70 &      115.5 &  0.66(8) &     36.9 &         \hdr{--\phantom{11.}}\\ 
    B1849+00     &      2180.20 &      787.0 &   2.2(2) &     64.1 &         \hdr{--\phantom{11.}}\\ 
    B1853+01     &       267.44 &       96.7 &  0.19(3) &     99.7 &      0.323 \\ 
    B1854+00     &       356.93 &       82.4 &   0.9(1) &    267.9 &      1.048 \\ 
    B1855+02     &       415.82 &      506.8 &   1.6(2) &    470.2 &      2.288 \\ 
    B1859+01     &       288.22 &      105.4 &  0.38(5) &     74.7 &      0.531 \\ 
    B1859+03     &       655.45 &      402.1 &   4.2(4) &   1061.3 &      3.498 \\ 
    B1859+07     &       644.00 &      252.8 &   0.9(1) &    339.1 &      1.830 \\ 
    B1900+01     &       729.30 &      245.2 &   5.5(6) &    106.5 &         \hdr{--\phantom{11.}}\\ 
    B1900+05     &       746.58 &      177.5 &   1.2(1) &    283.2 &      1.228 \\ 
    B1900+06     &       673.50 &      502.9 &   1.1(1) &     21.5 &         \hdr{--\phantom{11.}}\\ 
    B1901+10     &      1856.57 &      135.0 &  0.58(7) &    212.1 &      0.568 \\ 
    B1903+07     &       648.04 &      245.3 &   1.8(2) &     91.2 &      1.892 \\ 
    B1904+06     &       267.28 &      472.8 &   1.7(2) &     33.9 &         \hdr{--\phantom{11.}}\\ 
    B1906+09     &       830.27 &      249.8 &  0.23(3) &     17.7 &      0.127 \\ 
    B1907+02     &       989.83 &      171.7 &  0.63(7) &     37.7 &         \hdr{--\phantom{11.}}\\ 
    B1907+10     &       283.64 &      150.0 &   1.9(2) &    365.2 &      2.591 \\ 
    B1907+12     &      1441.74 &      258.6 &  0.28(4) &     28.2 &      0.196 \\ 
    B1910+10     &       409.35 &      147.0 &  0.22(3) &     47.1 &      0.196 \\ 
    B1911+09     &      1241.96 &      157.0 &  0.14(2) &     18.9 &      0.228 \\ 
    B1911+11     &       601.00 &      100.0 &  0.55(7) &     85.4 &      0.301 \\ 
    B1911+13     &       521.47 &      145.1 &   1.2(1) &     85.5 &      1.221 \\ 
    B1913+10     &       404.55 &      241.7 &   1.30(14) &    416.8 &      0.905 \\ 
    B1913+105    &       628.97 &      387.2 &  0.22(3) &     46.2 &      0.507 \\ 
    B1913+167    &      1616.23 &       62.6 &         \hdr{--\phantom{111.}}&     16.1 &         \hdr{--\phantom{11.}}\\ 
    B1914+09     &       270.25 &       61.0 &   0.9(1) &    298.6 &      0.721 \\ 
    B1914+13     &       281.84 &      237.0 &   1.2(1) &    616.7 &      2.043 \\ 
    B1915+13     &       194.63 &       94.5 &   1.9(2) &   1453.2 &      4.477 \\ 
    B1916+14     &      1181.02 &       27.2 &   1.0(1) &     14.3 &      0.362 \\ 
    B1919+14     &       618.18 &       91.6 &  0.68(8) &    217.6 &      1.060 \\ 
    B1921+17     &       547.21 &      142.5 &         \hdr{--\phantom{111.}}&    126.6 &      0.408 \\ 
    B1924+14     &      1324.92 &      211.4 &  0.48(6) &    126.6 &      0.860 \\ 
    B1924+16     &       579.82 &      176.9 &   1.3(2) &    179.1 &      0.735 \\ 
    B1925+18     &       482.77 &      254.0 &         \hdr{--\phantom{111.}}&    156.0 &      0.441 \\ 
    B1925+188    &       298.31 &       99.0 &         \hdr{--\phantom{111.}}&     77.3 &      0.385 \\ 
    B1929+15     &       314.36 &      140.0 &         \hdr{--\phantom{111.}}&     69.4 &      0.360 \\ 
    B1929+20     &       268.22 &      211.2 &   1.2(4) &    457.9 &      1.099 \\ 
    B1933+16     &       358.74 &      158.5 &  \hdr{42(6)} &     73.0 &         \hdr{--\phantom{11.}}\\ 
    B1933+17     &       654.41 &      214.6 &         \hdr{--\phantom{111.}} &     62.8 &      0.176 \\ 
    B1937+21     &         1.56 &       71.0 &  \hdr{13(5)} &    349.1 &     12.572 \\ 
    B1937+24     &       645.30 &      142.9 &         \hdr{--\phantom{111.}}&     39.4 &         \hdr{--\phantom{11.}}\\ 
    B1944+22     &      1334.45 &      140.0 &         \hdr{--\phantom{111.}}&     55.0 &      0.173 \\ 
    B2002+31     &      2111.26 &      234.8 &   1.8(1) &     68.2 &         \hdr{--\phantom{11.}}\\ 
    J0621+1002   &        28.85 &       36.6 &   1.9(3) &     11.4 &         \hdr{--\phantom{11.}}\\ 
    J0625+10     &       498.40 &       78.0 &         \hdr{--\phantom{111.}}&     14.5 &      0.086 \\ 
    J0631+1036   &       287.80 &      125.4 &         \hdr{--\phantom{111.}}&    175.3 &      0.941 \\ 
    J1829+0000   &       199.15 &      114.0 &         \hdr{--\phantom{111.}}&     52.4 &      0.370 \\ 
    J1843$-$0000 &       880.33 &      101.5 &   2.9(3) &     38.5 &         \hdr{--\phantom{11.}}\\ 
    J1844+00     &       460.50 &      345.5 &   8.6(9) &   1226.8 &      4.616 \\ 
    J1849+0127   &       542.16 &      207.3 &  0.46(9) &    143.2 &      0.444 \\ 
    J1849+0409   &       761.19 &       56.1 &         \hdr{--\phantom{111.}}&     29.0 &      0.312 \\ 
    J1851+0118   &       906.98 &      418.0 &  0.10(2) &     27.9 &      0.118 \\ 
    J1852+0305   &      1326.15 &      320.0 &   0.8(2) &     37.7 &      0.214 \\ 
    J1853+0056   &       275.58 &      180.9 &  0.21(4) &     55.3 &      0.281 \\ 
    J1853+0545   &       126.40 &      198.7 &   1.6(1.7) &      5.3 &         \hdr{--\phantom{11.}}\\ 
    J1854+0317   &      1366.45 &      404.0 &  0.12(1) &     34.9 &      0.153 \\ 
    J1855+0307   &       845.35 &      402.5 &   1.0(1) &    129.7 &      0.393 \\ 
    J1855+0422   &      1678.11 &      438.0 &  0.45(9) &    104.0 &      0.245 \\ 
    J1856+0102   &       620.22 &      554.0 &   0.4(1) &     66.3 &      0.195 \\ 
    J1856+0404   &       420.25 &      341.3 &  0.48(1) &     40.4 &      0.276 \\ 
    J1857+0143   &       139.76 &      249.0 &   0.7(2) &     37.2 &      0.486 \\ 
    J1857+0210   &       630.98 &      783.0 &  0.30(6) &     40.2 &      0.236 \\ 
    J1857+0526   &       349.95 &      466.4 &  0.66(8) &    145.5 &      0.645 \\ 
    J1858+0215   &       745.83 &      702.0 &  0.22(4) &     42.8 &      0.280 \\ 
    J1859+00     &       559.63 &      420.0 &   4.8(5) &    581.9 &     24.461 \\ 
    J1859+0601   &      1044.31 &      276.0 &  0.30(4) &     15.9 &      0.126 \\ 
    J1900+0227   &       374.26 &      201.1 &  0.33(7) &    111.6 &      0.414 \\ 
    J1901+00     &       777.66 &      345.5 &  0.35(4) &     32.4 &         \hdr{--\phantom{11.}}\\ 
    J1901+0254   &      1299.69 &      185.0 &  0.58(7) &    102.1 &      0.911 \\ 
    J1901+0320   &       636.58 &      393.0 &   0.9(1) &     67.3 &      0.301 \\ 
    J1901+0355   &       554.76 &      547.0 &  0.15(3) &     40.9 &      0.185 \\ 
    J1901+0413   &      2663.08 &      352.0 &   1.1(2) &    161.9 &      0.521 \\ 
    J1901+0435   &       690.58 &     1042.6 &         \hdr{--\phantom{111.}}&    106.9 &      4.244 \\ 
    J1901+0510   &       614.76 &      429.0 &  0.66(8) &     47.6 &      0.498 \\ 
    J1902+0248   &      1223.78 &      272.0 &  0.17(3) &     60.6 &      0.169 \\ 
    J1903+0601   &       374.12 &      388.0 &  0.26(4) &      9.7 &         \hdr{--\phantom{11.}}\\ 
    J1904+0412   &        71.09 &      185.9 &  0.23(5) &     68.4 &      0.271 \\ 
    J1904+0800   &       263.34 &      438.8 &  0.36(5) &     11.2 &      0.285 \\ 
    J1905+0600   &       441.21 &      730.1 &  0.42(5) &     85.6 &      0.401 \\ 
    J1905+0616   &       989.71 &      256.1 &  0.51(6) &     43.5 &      0.236 \\ 
    J1906+0912   &       775.34 &      265.0 &  0.32(6) &     34.0 &      0.149 \\ 
    J1907+0249   &       351.88 &      261.0 &   0.5(1) &    124.3 &      0.478 \\ 
    J1907+0345   &       240.15 &      311.7 &  0.17(3) &     21.5 &      0.133 \\ 
    J1907+0534   &      1138.40 &      524.0 &  0.36(7) &     24.6 &      0.096 \\ 
    J1907+0731   &       363.68 &      239.8 &  0.35(4) &     68.8 &      0.571 \\ 
    J1907+0740   &       574.70 &      332.0 &  0.41(8) &    121.4 &      0.327 \\ 
    J1907+0918   &       226.11 &      357.9 &  0.29(4) &    133.4 &      0.263 \\ 
    J1907+1149   &      1420.16 &      202.8 &         \hdr{--\phantom{111.}}&     30.4 &      0.156 \\ 
    J1908+0457   &       846.79 &      360.0 &   0.9(1) &    274.4 &      0.958 \\ 
    J1908+0500   &       291.02 &      201.4 &  0.79(9) &     48.5 &         \hdr{--\phantom{11.}}\\ 
    J1908+0734   &       212.35 &       11.1 &  0.54(6) &     36.0 &      0.205 \\ 
    J1908+0839   &       185.40 &      512.1 &  0.49(1) &    114.4 &      0.403 \\ 
    J1908+0909   &       336.55 &      467.5 &  0.22(4) &    110.7 &      0.340 \\ 
    J1909+0616   &       755.99 &      352.0 &  0.33(7) &     10.3 &         \hdr{--\phantom{111.}}\\ 
    J1909+0912   &       222.95 &      421.5 &  0.35(7) &    125.8 &      0.533 \\ 
    J1910+0534   &       452.87 &      484.0 &  0.41(8) &     62.4 &      0.444 \\ 
    J1910+0714   &      2712.42 &      124.1 &  0.36(5) &    137.3 &      0.287 \\ 
    J1910+0728   &       325.42 &      283.7 &   0.8(1) &    189.8 &      0.887 \\ 
    J1910+1256   &         4.98 &       38.1 &   0.5(1) &    139.7 &      0.497 \\ 
    J1913+0832   &       134.41 &      355.2 &   0.6(1) &    187.9 &      0.999 \\ 
    J1913+0904   &       163.25 &       95.3 &         \hdr{--\phantom{111.}}&     96.7 &      0.224 \\ 
    J1913+1000   &       837.15 &      422.0 &  0.53(6) &     28.8 &      0.522 \\ 
    J1913+1011   &        35.91 &      178.8 &   0.5(1) &    111.0 &      0.434 \\ 
    J1913+1145   &       306.07 &      637.0 &  0.43(9) &    126.5 &      0.403 \\ 
    J1913+1330   &       923.39 &      175.6 &         \hdr{--\phantom{111.}}&    213.6 &         \hdr{--\phantom{11.}}\\ 
    J1914+0631   &       693.81 &       58.0 &   0.3(1) &     36.9 &      0.140 \\ 
    J1915+0738   &      1542.70 &       39.0 &  0.34(4) &    109.1 &      0.254 \\ 
    J1915+0752   &      2058.31 &      105.3 &  0.21(3) &     18.2 &      0.238 \\ 
    J1915+0838   &       342.78 &      358.0 &  0.29(4) &     12.3 &         \hdr{--\phantom{11.}}\\ 
    J1915+1410   &       297.49 &      273.7 &         \hdr{--\phantom{111.}}&     11.6 &      0.134 \\ 
    J1916+0748   &       541.75 &      304.0 &   2.8(3) &     66.8 &         \hdr{--\phantom{11.}}\\ 
    J1916+0844   &       440.00 &      339.4 &  0.44(5) &     89.9 &      0.526 \\ 
    J1916+0852   &      2182.75 &      295.0 &  0.13(2) &     36.6 &      0.148 \\ 
    J1920+1040   &      2215.80 &      304.0 &  0.57(7) &     24.5 &      0.092 \\ 
    J1920+1110   &       509.89 &      182.0 &  0.39(8) &     22.9 &      0.288 \\ 
    J1921+1544   &       143.58 &      385.0 &         \hdr{--\phantom{111.}}&     65.5 &      0.211 \\ 
    J1922+1733   &       236.17 &      238.0 &         \hdr{--\phantom{111.}}&    435.6 &      1.157 \\ 
    J1924+1639   &       158.04 &      208.0 &         \hdr{--\phantom{111.}}&     73.6 &      0.207 \\ 
    J1926+2016   &       299.07 &      247.0 &         \hdr{--\phantom{111.}}&     12.0 &      0.122 \\ 
    J1928+1923   &       817.33 &      476.0 &         \hdr{--\phantom{111.}}&    221.7 &      0.639 \\ 
    J1929+1955   &       257.83 &      281.0 &         \hdr{--\phantom{111.}}&     25.1 &      0.421 \\ 
    J1930+17     &      1609.69 &      201.0 &         \hdr{--\phantom{111.}}&     30.9 &         \hdr{--\phantom{11.}}\\ 
    J1931+1952   &       501.12 &      441.0 &         \hdr{--\phantom{111.}}&     71.9 &      0.126 \\ 
    J1935+2025   &        80.12 &      182.0 &         \hdr{--\phantom{111.}}&     79.6 &      0.527 \\ 
    J1936+21     &       642.93 &      264.0 &         \hdr{--\phantom{111.}}&     13.6 &         \hdr{--\phantom{11.}}\\ 
    J1938+2213   &       166.12 &       91.0 &         \hdr{--\phantom{111.}}&     20.4 &         \hdr{--\phantom{11.}}\\ 
    J1946+2611   &       435.06 &      165.0 &         \hdr{--\phantom{111.}}&    232.0 &      0.697 \\ 
    J1957+2831   &       307.68 &      139.0 &   1.0(2) &     34.4 &         \hdr{--\phantom{11.}}\\ 
\end{longtable}
\end{center}
\vspace{-5mm}
{\footnotesize \textsc{Note.} --- Values for period, DM, and ``ATNF $S_{1400}$'' are taken from the ATNF Catalogue \citep{mhth05}\\
}

\begin{table}
\centering
    \caption{Synthetic pulsar signal parameters \label{tab:injparams}}
\begin{tabular}{lcccccc}
    \hline\hline\\[-2mm]
    Parameter & \multicolumn{6}{c}{Possible values}\\
    \hline\\[-2mm]
                   & 0.766 & 1.102 & 2.218 & 5.218 & 10.870 & 18.505 \\
        Period, ms & 26.965 & 61.631 & 126.175 & 286.555 & 533.320 & 850.158 \\
                   & 1657.496 & 2643.410 & 3927.013 & 5580.899 & 10964.532 & \\
        DM, \dmunit & 10 & 40 & 150 & 325 & 400 & 600 \\
        FWHM, \% phase & 1.5 & 2.6 & 5.9 & 11.9 & 24.3 & \\
    \hline\\[-2mm]
\end{tabular}
\end{table}

\clearpage

\begin{thebibliography}{}
\expandafter\ifx\csname natexlab\endcsname\relax\def\natexlab#1{#1}\fi

\bibitem[{{Allen} {et~al.}(2013){Allen}, {Knispel}, {Cordes}, {Deneva},
  {Hessels}, {Anderson}, {Aulbert}, {Bock}, {Brazier}, {Chatterjee},
  {Demorest}, {Eggenstein}, {Fehrmann}, {Gotthelf}, {Hammer}, {Kaspi},
  {Kramer}, {Lyne}, {Machenschalk}, {McLaughlin}, {Messenger}, {Pletsch},
  {Ransom}, {Stairs}, {Stappers}, {Bhat}, {Bogdanov}, {Camilo}, {Champion},
  {Crawford}, {Desvignes}, {Freire}, {Heald}, {Jenet}, {Lazarus}, {Lee}, {van
  Leeuwen}, {Lynch}, {Papa}, {Prix}, {Rosen}, {Scholz}, {Siemens}, {Stovall},
  {Venkataraman}, \& {Zhu}}]{akc+13}
{Allen}, B., {Knispel}, B., {Cordes}, J.~M., {et~al.} 2013, \apj, 773, 91

\bibitem[{{Alpar} {et~al.}(1982){Alpar}, {Cheng}, {Ruderman}, \&
  {Shaham}}]{acrs82}
{Alpar}, M.~A., {Cheng}, A.~F., {Ruderman}, M.~A., \& {Shaham}, J. 1982, \nat,
  300, 728

\bibitem[{{Antoniadis} {et~al.}(2013){Antoniadis}, {Freire}, {Wex}, {Tauris},
  {Lynch}, {van Kerkwijk}, {Kramer}, {Bassa}, {Dhillon}, {Driebe}, {Hessels},
  {Kaspi}, {Kondratiev}, {Langer}, {Marsh}, {McLaughlin}, {Pennucci}, {Ransom},
  {Stairs}, {van Leeuwen}, {Verbiest}, \& {Whelan}}]{afw+13}
{Antoniadis}, J., {Freire}, P.~C.~C., {Wex}, N., {et~al.} 2013, Science, 340,
  448

\bibitem[{{Barr} {et~al.}(2013){Barr}, {Champion}, {Kramer}, {Eatough},
  {Freire}, {Karuppusamy}, {Lee}, {Verbiest}, {Bassa}, {Lyne}, {Stappers},
  {Lorimer}, \& {Klein}}]{bck+13}
{Barr}, E.~D., {Champion}, D.~J., {Kramer}, M., {et~al.} 2013, \mnras, 435,
  2234

\bibitem[{{Bates} {et~al.}(2014){Bates}, {Lorimer}, {Rane}, \&
  {Swiggum}}]{blrs14}
{Bates}, S.~D., {Lorimer}, D.~R., {Rane}, A., \& {Swiggum}, J. 2014, \mnras,
  439, 2893

\bibitem[{{Burke-Spolaor} {et~al.}(2011){Burke-Spolaor}, {Bailes}, {Ekers},
  {Macquart}, \& {Crawford}}]{bbe+11}
{Burke-Spolaor}, S., {Bailes}, M., {Ekers}, R., {Macquart}, J.-P., \&
  {Crawford}, III, F. 2011, \apj, 727, 18

\bibitem[{{Camilo} {et~al.}(2007){Camilo}, {Ransom}, {Halpern}, \&
  {Reynolds}}]{crhr07}
{Camilo}, F., {Ransom}, S.~M., {Halpern}, J.~P., \& {Reynolds}, J. 2007, \apjl,
  666, L93

\bibitem[{{Camilo} {et~al.}(2006){Camilo}, {Ransom}, {Halpern}, {Reynolds},
  {Helfand}, {Zimmerman}, \& {Sarkissian}}]{crh+06}
{Camilo}, F., {Ransom}, S.~M., {Halpern}, J.~P., {et~al.} 2006, \nat, 442, 892

\bibitem[{{Champion} {et~al.}(2008){Champion}, {Ransom}, {Lazarus}, {Camilo},
  {Bassa}, {Kaspi}, {Nice}, {Freire}, {Stairs}, {van Leeuwen}, {Stappers},
  {Cordes}, {Hessels}, {Lorimer}, {Arzoumanian}, {Backer}, {Bhat},
  {Chatterjee}, {Cognard}, {Deneva}, {Faucher-Gigu{\`e}re}, {Gaensler}, {Han},
  {Jenet}, {Kasian}, {Kondratiev}, {Kramer}, {Lazio}, {McLaughlin},
  {Venkataraman}, \& {Vlemmings}}]{crl+08}
{Champion}, D.~J., {Ransom}, S.~M., {Lazarus}, P., {et~al.} 2008, Science, 320,
  1309

\bibitem[{{Coenen} {et~al.}(2014){Coenen}, {van Leeuwen}, {Hessels},
  {Stappers}, {Kondratiev}, {Alexov}, {Breton}, {Bilous}, {Cooper}, {Falcke},
  {Fallows}, {Gajjar}, {Grie{\ss}meier}, {Hassall}, {Karastergiou}, {Keane},
  {Kramer}, {Kuniyoshi}, {Noutsos}, {Os{\l}owski}, {Pilia}, {Serylak},
  {Schrijvers}, {Sobey}, {ter Veen}, {Verbiest}, {Weltevrede}, {Wijnholds},
  {Zagkouris}, {van Amesfoort}, {Anderson}, {Asgekar}, {Avruch}, {Bell},
  {Bentum}, {Bernardi}, {Best}, {Bonafede}, {Breitling}, {Broderick},
  {Br{\"u}ggen}, {Butcher}, {Ciardi}, {Corstanje}, {Deller}, {Duscha},
  {Eisl{\"o}ffel}, {Fender}, {Ferrari}, {Frieswijk}, {Garrett}, {de Gasperin},
  {de Geus}, {Gunst}, {Hamaker}, {Heald}, {Hoeft}, {van der Horst}, {Juette},
  {Kuper}, {Law}, {Mann}, {McFadden}, {McKay-Bukowski}, {McKean}, {Munk},
  {Orru}, {Paas}, {Pandey-Pommier}, {Polatidis}, {Reich}, {Renting},
  {R{\"o}ttgering}, {Rowlinson}, {Scaife}, {Schwarz}, {Sluman}, {Smirnov},
  {Swinbank}, {Tagger}, {Tang}, {Tasse}, {Thoudam}, {Toribio}, {Vermeulen},
  {Vocks}, {van Weeren}, {Wucknitz}, {Zarka}, \& {Zensus}}]{cvh+14}
{Coenen}, T., {van Leeuwen}, J., {Hessels}, J.~W.~T., {et~al.} 2014, ArXiv
  e-prints, arXiv:1408.0411

\bibitem[{{Cordes}(2002)}]{cor02}
{Cordes}, J.~M. 2002, in Astronomical Society of the Pacific Conference Series,
  Vol. 278, Single-Dish Radio Astronomy: Techniques and Applications, ed.
  S.~{Stanimirovic}, D.~{Altschuler}, P.~{Goldsmith}, \& C.~{Salter}, 227--250

\bibitem[{{Cordes} \& {Chernoff}(1997)}]{cc97}
{Cordes}, J.~M., \& {Chernoff}, D.~F. 1997, \apj, 482, 971

\bibitem[{{Cordes} \& {Lazio}(2002)}]{cl02}
{Cordes}, J.~M., \& {Lazio}, T.~J.~W. 2002, ArXiv Astrophysics e-prints,
  astro-ph/0207156

\bibitem[{{Cordes} \& {McLaughlin}(2003)}]{cm03}
{Cordes}, J.~M., \& {McLaughlin}, M.~A. 2003, \apj, 596, 1142

\bibitem[{{Cordes} {et~al.}(2006){Cordes}, {Freire}, {Lorimer}, {Camilo},
  {Champion}, {Nice}, {Ramachandran}, {Hessels}, {Vlemmings}, {van Leeuwen},
  {Ransom}, {Bhat}, {Arzoumanian}, {McLaughlin}, {Kaspi}, {Kasian}, {Deneva},
  {Reid}, {Chatterjee}, {Han}, {Backer}, {Stairs}, {Deshpande}, \&
  {Faucher-Gigu{\`e}re}}]{cfl+06}
{Cordes}, J.~M., {Freire}, P.~C.~C., {Lorimer}, D.~R., {et~al.} 2006, \apj,
  637, 446

\bibitem[{{Crawford} {et~al.}(2012){Crawford}, {Stovall}, {Lyne}, {Stappers},
  {Nice}, {Stairs}, {Lazarus}, {Hessels}, {Freire}, {Allen}, {Bhat},
  {Bogdanov}, {Brazier}, {Camilo}, {Champion}, {Chatterjee}, {Cognard},
  {Cordes}, {Deneva}, {Desvignes}, {Jenet}, {Kaspi}, {Knispel}, {Kramer}, {van
  Leeuwen}, {Lorimer}, {Lynch}, {McLaughlin}, {Ransom}, {Scholz}, {Siemens}, \&
  {Venkataraman}}]{csl+12}
{Crawford}, F., {Stovall}, K., {Lyne}, A.~G., {et~al.} 2012, \apj, 757, 90

\bibitem[{{Demorest} {et~al.}(2010){Demorest}, {Pennucci}, {Ransom}, {Roberts},
  \& {Hessels}}]{dpr+10}
{Demorest}, P.~B., {Pennucci}, T., {Ransom}, S.~M., {Roberts}, M.~S.~E., \&
  {Hessels}, J.~W.~T. 2010, \nat, 467, 1081

\bibitem[{{Deneva} {et~al.}(2013){Deneva}, {Stovall}, {McLaughlin}, {Bates},
  {Freire}, {Martinez}, {Jenet}, \& {Bagchi}}]{dsm+13}
{Deneva}, J.~S., {Stovall}, K., {McLaughlin}, M.~A., {et~al.} 2013, \apj, 775,
  51

\bibitem[{{Deneva} {et~al.}(2009){Deneva}, {Cordes}, {McLaughlin}, {Nice},
  {Lorimer}, {Crawford}, {Bhat}, {Camilo}, {Champion}, {Freire}, {Edel},
  {Kondratiev}, {Hessels}, {Jenet}, {Kasian}, {Kaspi}, {Kramer}, {Lazarus},
  {Ransom}, {Stairs}, {Stappers}, {van Leeuwen}, {Brazier}, {Venkataraman},
  {Zollweg}, \& {Bogdanov}}]{dcm+09}
{Deneva}, J.~S., {Cordes}, J.~M., {McLaughlin}, M.~A., {et~al.} 2009, \apj,
  703, 2259

\bibitem[{{Desvignes} {et~al.}(2013){Desvignes}, {Cognard}, {Champion},
  {Lazarus}, {Lespagnol}, {Smith}, \& {Theureau}}]{dcc+13}
{Desvignes}, G., {Cognard}, I., {Champion}, D., {et~al.} 2013, in IAU
  Symposium, Vol. 291, IAU Symposium, ed. J.~{van Leeuwen}, 375--377

\bibitem[{{Dewey} {et~al.}(1985){Dewey}, {Taylor}, {Weisberg}, \&
  {Stokes}}]{dtws85}
{Dewey}, R.~J., {Taylor}, J.~H., {Weisberg}, J.~M., \& {Stokes}, G.~H. 1985,
  \apjl, 294, L25

\bibitem[{{Dowd} {et~al.}(2000){Dowd}, {Sisk}, \& {Hagen}}]{dsh00}
{Dowd}, A., {Sisk}, W., \& {Hagen}, J. 2000, in Astronomical Society of the
  Pacific Conference Series, Vol. 202, IAU Colloq. 177: Pulsar Astronomy - 2000
  and Beyond, ed. M.~{Kramer}, N.~{Wex}, \& R.~{Wielebinski}, 275

\bibitem[{{Eatough} {et~al.}(2009){Eatough}, {Keane}, \& {Lyne}}]{ekl09}
{Eatough}, R.~P., {Keane}, E.~F., \& {Lyne}, A.~G. 2009, \mnras, 395, 410

\bibitem[{{Eatough} {et~al.}(2013){Eatough}, {Falcke}, {Karuppusamy}, {Lee},
  {Champion}, {Keane}, {Desvignes}, {Schnitzeler}, {Spitler}, {Kramer},
  {Klein}, {Bassa}, {Bower}, {Brunthaler}, {Cognard}, {Deller}, {Demorest},
  {Freire}, {Kraus}, {Lyne}, {Noutsos}, {Stappers}, \& {Wex}}]{efk+13}
{Eatough}, R.~P., {Falcke}, H., {Karuppusamy}, R., {et~al.} 2013, \nat, 501,
  391

\bibitem[{{Faucher-Gigu{\`e}re} \& {Kaspi}(2006)}]{fk06}
{Faucher-Gigu{\`e}re}, C.-A., \& {Kaspi}, V.~M. 2006, \apj, 643, 332

\bibitem[{{Hankins} {et~al.}(2003){Hankins}, {Kern}, {Weatherall}, \&
  {Eilek}}]{hkwe03}
{Hankins}, T.~H., {Kern}, J.~S., {Weatherall}, J.~C., \& {Eilek}, J.~A. 2003,
  \nat, 422, 141

\bibitem[{{Haslam} {et~al.}(1982){Haslam}, {Salter}, {Stoffel}, \&
  {Wilson}}]{hssw82}
{Haslam}, C.~G.~T., {Salter}, C.~J., {Stoffel}, H., \& {Wilson}, W.~E. 1982,
  \aaps, 47, 1

\bibitem[{{Henning} {et~al.}(2010){Henning}, {Springob}, {Minchin}, {Momjian},
  {Catinella}, {McIntyre}, {Day}, {Muller}, {Koribalski}, {Rosenberg},
  {Schneider}, {Staveley-Smith}, \& {van Driel}}]{hsb+10}
{Henning}, P.~A., {Springob}, C.~M., {Minchin}, R.~F., {et~al.} 2010, \aj, 139,
  2130

\bibitem[{{Hermsen} {et~al.}(2013){Hermsen}, {Hessels}, {Kuiper}, {van
  Leeuwen}, {Mitra}, {de Plaa}, {Rankin}, {Stappers}, {Wright}, {Basu},
  {Alexov}, {Coenen}, {Grie{\ss}meier}, {Hassall}, {Karastergiou}, {Keane},
  {Kondratiev}, {Kramer}, {Kuniyoshi}, {Noutsos}, {Serylak}, {Pilia}, {Sobey},
  {Weltevrede}, {Zagkouris}, {Asgekar}, {Avruch}, {Batejat}, {Bell}, {Bell},
  {Bentum}, {Bernardi}, {Best}, {B{\^i}rzan}, {Bonafede}, {Breitling},
  {Broderick}, {Br{\"u}ggen}, {Butcher}, {Ciardi}, {Duscha}, {Eisl{\"o}ffel},
  {Falcke}, {Fender}, {Ferrari}, {Frieswijk}, {Garrett}, {de Gasperin}, {de
  Geus}, {Gunst}, {Heald}, {Hoeft}, {Horneffer}, {Iacobelli}, {Kuper}, {Maat},
  {Macario}, {Markoff}, {McKean}, {Mevius}, {Miller-Jones}, {Morganti}, {Munk},
  {Orr{\'u}}, {Paas}, {Pandey-Pommier}, {Pandey}, {Pizzo}, {Polatidis},
  {Rawlings}, {Reich}, {R{\"o}ttgering}, {Scaife}, {Schoenmakers}, {Shulevski},
  {Sluman}, {Steinmetz}, {Tagger}, {Tang}, {Tasse}, {ter Veen}, {Vermeulen},
  {van de Brink}, {van Weeren}, {Wijers}, {Wise}, {Wucknitz}, {Yatawatta}, \&
  {Zarka}}]{hhk+13}
{Hermsen}, W., {Hessels}, J.~W.~T., {Kuiper}, L., {et~al.} 2013, Science, 339,
  436

\bibitem[{{Hessels} {et~al.}(2006){Hessels}, {Ransom}, {Stairs}, {Freire},
  {Kaspi}, \& {Camilo}}]{hrs+06}
{Hessels}, J.~W.~T., {Ransom}, S.~M., {Stairs}, I.~H., {et~al.} 2006, Science,
  311, 1901

\bibitem[{{Hessels} {et~al.}(2008){Hessels}, {Nice}, {Gaensler}, {Kaspi},
  {Lorimer}, {Champion}, {Lyne}, {Kramer}, {Cordes}, {Freire}, {Camilo},
  {Ransom}, {Deneva}, {Bhat}, {Cognard}, {Crawford}, {Jenet}, {Kasian},
  {Lazarus}, {van Leeuwen}, {McLaughlin}, {Stairs}, {Stappers}, \&
  {Venkataraman}}]{hng+08}
{Hessels}, J.~W.~T., {Nice}, D.~J., {Gaensler}, B.~M., {et~al.} 2008, \apjl,
  682, L41

\bibitem[{{Hotan} {et~al.}(2004){Hotan}, {van Straten}, \&
  {Manchester}}]{hvm04}
{Hotan}, A.~W., {van Straten}, W., \& {Manchester}, R.~N. 2004, \pasa, 21, 302

\bibitem[{{Karako-Argaman} {et~al.}(2015){Karako-Argaman}, {Kaspi}, {Lynch},
  {Hessels}, {Kondratiev}, {McLaughlin}, {Ransom}, {Archibald}, {Boyles},
  {Jenet}, {Kaplan}, {Levin}, {Lorimer}, {Madsen}, {Roberts}, {Siemens},
  {Stairs}, {Stovall}, {Swiggum}, \& {van Leeuwen}}]{kkl+15}
{Karako-Argaman}, C., {Kaspi}, V.~M., {Lynch}, R.~S., {et~al.} 2015, ArXiv
  e-prints, arXiv:1503.05170

\bibitem[{{Keith} {et~al.}(2010){Keith}, {Jameson}, {van Straten}, {Bailes},
  {Johnston}, {Kramer}, {Possenti}, {Bates}, {Bhat}, {Burgay}, {Burke-Spolaor},
  {D'Amico}, {Levin}, {McMahon}, {Milia}, \& {Stappers}}]{kjs+10}
{Keith}, M.~J., {Jameson}, A., {van Straten}, W., {et~al.} 2010, \mnras, 409,
  619

\bibitem[{{Kiddle} {et~al.}(2011){Kiddle}, {Andrecut}, {Brazier}, {Chatterjee},
  {Chen}, {Cordes}, {Curry}, {Este}, {Eymere}, {Federl}, {Fong}, {Grimstrup},
  {Guram}, {Kaspi}, {Klodzinski}, {Lazarus}, {Mahadevan}, {Mourad}, {Mourad},
  {Pragides}, {Rosolowsky}, {Said}, {Samoilov}, {Smith}, {Stairs}, {Tan},
  {Tan}, {Taylor}, \& {Willis}}]{kab+11}
{Kiddle}, C., {Andrecut}, M., {Brazier}, A., {et~al.} 2011, in Astronomical
  Society of the Pacific Conference Series, Vol. 442, Astronomical Data
  Analysis Software and Systems XX, ed. I.~N. {Evans}, A.~{Accomazzi}, D.~J.
  {Mink}, \& A.~H. {Rots}, 669

\bibitem[{{Kondratiev} {et~al.}(2009){Kondratiev}, {McLaughlin}, {Lorimer},
  {Burgay}, {Possenti}, {Turolla}, {Popov}, \& {Zane}}]{kml+09}
{Kondratiev}, V.~I., {McLaughlin}, M.~A., {Lorimer}, D.~R., {et~al.} 2009,
  \apj, 702, 692

\bibitem[{{Kramer} {et~al.}(2006{\natexlab{a}}){Kramer}, {Lyne}, {O'Brien},
  {Jordan}, \& {Lorimer}}]{klo+06}
{Kramer}, M., {Lyne}, A.~G., {O'Brien}, J.~T., {Jordan}, C.~A., \& {Lorimer},
  D.~R. 2006{\natexlab{a}}, Science, 312, 549

\bibitem[{{Kramer} {et~al.}(2006{\natexlab{b}}){Kramer}, {Stairs},
  {Manchester}, {McLaughlin}, {Lyne}, {Ferdman}, {Burgay}, {Lorimer},
  {Possenti}, {D'Amico}, {Sarkissian}, {Hobbs}, {Reynolds}, {Freire}, \&
  {Camilo}}]{ksm+06}
{Kramer}, M., {Stairs}, I.~H., {Manchester}, R.~N., {et~al.}
  2006{\natexlab{b}}, Science, 314, 97

\bibitem[{{Levin} {et~al.}(2010){Levin}, {Bailes}, {Bates}, {Bhat}, {Burgay},
  {Burke-Spolaor}, {D'Amico}, {Johnston}, {Keith}, {Kramer}, {Milia},
  {Possenti}, {Rea}, {Stappers}, \& {van Straten}}]{lbb+10}
{Levin}, L., {Bailes}, M., {Bates}, S., {et~al.} 2010, \apjl, 721, L33

\bibitem[{{Liu} {et~al.}(2013){Liu}, {McIntyre}, {Terzian}, {Minchin},
  {Anderson}, {Churchwell}, {Lebron}, \& {Anish Roshi}}]{lmt+13}
{Liu}, B., {McIntyre}, T., {Terzian}, Y., {et~al.} 2013, \aj, 146, 80

\bibitem[{{Lorimer} \& {Kramer}(2004)}]{lk04}
{Lorimer}, D.~R., \& {Kramer}, M. 2004, {Handbook of Pulsar Astronomy}, ed.
  R.~{Ellis}, J.~{Huchra}, S.~{Kahn}, G.~{Rieke}, \& P.~B. {Stetson}

\bibitem[{{Lorimer} {et~al.}(2006{\natexlab{a}}){Lorimer}, {Stairs}, {Freire},
  {Cordes}, {Camilo}, {Faulkner}, {Lyne}, {Nice}, {Ransom}, {Arzoumanian},
  {Manchester}, {Champion}, {van Leeuwen}, {Mclaughlin}, {Ramachandran},
  {Hessels}, {Vlemmings}, {Deshpande}, {Bhat}, {Chatterjee}, {Han}, {Gaensler},
  {Kasian}, {Deneva}, {Reid}, {Lazio}, {Kaspi}, {Crawford}, {Lommen}, {Backer},
  {Kramer}, {Stappers}, {Hobbs}, {Possenti}, {D'Amico}, \& {Burgay}}]{lsf+06}
{Lorimer}, D.~R., {Stairs}, I.~H., {Freire}, P.~C., {et~al.}
  2006{\natexlab{a}}, \apj, 640, 428

\bibitem[{{Lorimer} {et~al.}(2006{\natexlab{b}}){Lorimer}, {Faulkner}, {Lyne},
  {Manchester}, {Kramer}, {McLaughlin}, {Hobbs}, {Possenti}, {Stairs},
  {Camilo}, {Burgay}, {D'Amico}, {Corongiu}, \& {Crawford}}]{lfl+06}
{Lorimer}, D.~R., {Faulkner}, A.~J., {Lyne}, A.~G., {et~al.}
  2006{\natexlab{b}}, \mnras, 372, 777

\bibitem[{{Lyne} {et~al.}(2010){Lyne}, {Hobbs}, {Kramer}, {Stairs}, \&
  {Stappers}}]{lhk+10}
{Lyne}, A., {Hobbs}, G., {Kramer}, M., {Stairs}, I., \& {Stappers}, B. 2010,
  Science, 329, 408

\bibitem[{{Manchester} {et~al.}(2005){Manchester}, {Hobbs}, {Teoh}, \&
  {Hobbs}}]{mhth05}
{Manchester}, R.~N., {Hobbs}, G.~B., {Teoh}, A., \& {Hobbs}, M. 2005, \aj, 129,
  1993

\bibitem[{{Manchester} {et~al.}(2001){Manchester}, {Lyne}, {Camilo}, {Bell},
  {Kaspi}, {D'Amico}, {McKay}, {Crawford}, {Stairs}, {Possenti}, {Kramer}, \&
  {Sheppard}}]{mlc+01}
{Manchester}, R.~N., {Lyne}, A.~G., {Camilo}, F., {et~al.} 2001, \mnras, 328,
  17

\bibitem[{{Olausen} \& {Kaspi}(2014)}]{ok14}
{Olausen}, S.~A., \& {Kaspi}, V.~M. 2014, \apjs, 212, 6

\bibitem[{{Platania} {et~al.}(1998){Platania}, {Bensadoun}, {Bersanelli}, {De
  Amici}, {Kogut}, {Levin}, {Maino}, \& {Smoot}}]{pbb+98}
{Platania}, P., {Bensadoun}, M., {Bersanelli}, M., {et~al.} 1998, \apj, 505,
  473

\bibitem[{{Ransom}(2001)}]{ran01}
{Ransom}, S.~M. 2001, PhD thesis, Harvard University

\bibitem[{{Ransom} {et~al.}(2002){Ransom}, {Eikenberry}, \&
  {Middleditch}}]{rem02}
{Ransom}, S.~M., {Eikenberry}, S.~S., \& {Middleditch}, J. 2002, \aj, 124, 1788

\bibitem[{{Scholz} {et~al.}(2015){Scholz}, {Kaspi}, {Lyne}, {Stappers},
  {Bogdanov}, {Cordes}, {Crawford}, {Ferdman}, {Freire}, {Hessels}, {Lorimer},
  {Stairs}, {Allen}, {Brazier}, {Camilo}, {Cardoso}, {Chatterjee}, {Deneva},
  {Jenet}, {Karako-Argaman}, {Knispel}, {Lazarus}, {Lee}, {van Leeuwen},
  {Lynch}, {Madsen}, {McLaughlin}, {Ransom}, {Siemens}, {Spitler}, {Stovall},
  {Swiggum}, {Venkataraman}, \& {Zhu}}]{skl+15}
{Scholz}, P., {Kaspi}, V.~M., {Lyne}, A.~G., {et~al.} 2015, ArXiv e-prints,
  arXiv:1501.03746

\bibitem[{{Shannon} \& {Johnston}(2013)}]{sj13}
{Shannon}, R.~M., \& {Johnston}, S. 2013, \mnras, 435, L29

\bibitem[{{Spitler}(2013)}]{spi13}
{Spitler}, L.~G. 2013, PhD thesis, Cornell University

\bibitem[{{Spitler} {et~al.}(2014){Spitler}, {Cordes}, {Hessels}, {Lorimer},
  {McLaughlin}, {Chatterjee}, {Crawford}, {Deneva}, {Kaspi}, {Wharton},
  {Allen}, {Bogdanov}, {Brazier}, {Camilo}, {Freire}, {Jenet},
  {Karako-Argaman}, {Knispel}, {Lazarus}, {Lee}, {van Leeuwen}, {Lynch},
  {Ransom}, {Scholz}, {Siemens}, {Stairs}, {Stovall}, {Swiggum},
  {Venkataraman}, {Zhu}, {Aulbert}, \& {Fehrmann}}]{sch+14}
{Spitler}, L.~G., {Cordes}, J.~M., {Hessels}, J.~W.~T., {et~al.} 2014, \apj,
  790, 101

\bibitem[{{Stovall}(2013)}]{sto13}
{Stovall}, K. 2013, PhD thesis, The University of Texas at San Antonio

\bibitem[{{Stovall} {et~al.}(2014){Stovall}, {Lynch}, {Ransom}, {Archibald},
  {Banaszak}, {Biwer}, {Boyles}, {Dartez}, {Day}, {Ford}, {Flanigan}, {Garcia},
  {Hessels}, {Hinojosa}, {Jenet}, {Kaplan}, {Karako-Argaman}, {Kaspi},
  {Kondratiev}, {Leake}, {Lorimer}, {Lunsford}, {Martinez}, {Mata},
  {McLaughlin}, {Roberts}, {Rohr}, {Siemens}, {Stairs}, {van Leeuwen},
  {Walker}, \& {Wells}}]{slr+14}
{Stovall}, K., {Lynch}, R.~S., {Ransom}, S.~M., {et~al.} 2014, \apj, 791, 67

\bibitem[{{Swiggum} {et~al.}(2014){Swiggum}, {Lorimer}, {McLaughlin}, {Bates},
  {Champion}, {Ransom}, {Lazarus}, {Brazier}, {Hessels}, {Nice}, {Ellis},
  {Senty}, {Allen}, {Bhat}, {Bogdanov}, {Camilo}, {Chatterjee}, {Cordes},
  {Crawford}, {Deneva}, {Freire}, {Jenet}, {Karako-Argaman}, {Kaspi},
  {Knispel}, {Lee}, {van Leeuwen}, {Lynch}, {Lyne}, {Scholz}, {Siemens},
  {Stairs}, {Stappers}, {Stovall}, {Venkataraman}, \& {Zhu}}]{slm+14}
{Swiggum}, J.~K., {Lorimer}, D.~R., {McLaughlin}, M.~A., {et~al.} 2014, \apj,
  787, 137

\bibitem[{{Thornton} {et~al.}(2013){Thornton}, {Stappers}, {Bailes},
  {Barsdell}, {Bates}, {Bhat}, {Burgay}, {Burke-Spolaor}, {Champion}, {Coster},
  {D'Amico}, {Jameson}, {Johnston}, {Keith}, {Kramer}, {Levin}, {Milia}, {Ng},
  {Possenti}, \& {van Straten}}]{tsb+13}
{Thornton}, D., {Stappers}, B., {Bailes}, M., {et~al.} 2013, Science, 341, 53

\bibitem[{{Young} {et~al.}(1999){Young}, {Manchester}, \& {Johnston}}]{ymj99}
{Young}, M.~D., {Manchester}, R.~N., \& {Johnston}, S. 1999, \nat, 400, 848

\bibitem[{{Zhang} {et~al.}(2000){Zhang}, {Harding}, \& {Muslimov}}]{zhm00}
{Zhang}, B., {Harding}, A.~K., \& {Muslimov}, A.~G. 2000, \apjl, 531, L135

\bibitem[{{Zhu} {et~al.}(2014){Zhu}, {Berndsen}, {Madsen}, {Tan}, {Stairs},
  {Brazier}, {Lazarus}, {Lynch}, {Scholz}, {Stovall}, {Ransom}, {Banaszak},
  {Biwer}, {Cohen}, {Dartez}, {Flanigan}, {Lunsford}, {Martinez}, {Mata},
  {Rohr}, {Walker}, {Allen}, {Bhat}, {Bogdanov}, {Camilo}, {Chatterjee},
  {Cordes}, {Crawford}, {Deneva}, {Desvignes}, {Ferdman}, {Freire}, {Hessels},
  {Jenet}, {Kaplan}, {Kaspi}, {Knispel}, {Lee}, {van Leeuwen}, {Lyne},
  {McLaughlin}, {Siemens}, {Spitler}, \& {Venkataraman}}]{zbm+14}
{Zhu}, W.~W., {Berndsen}, A., {Madsen}, E.~C., {et~al.} 2014, \apj, 781, 117

\end{thebibliography}
\end{document}